\documentclass[aps,prd,twocolumn,floats,floatfix,nofootinbib,superscriptaddress,longbibliography]{revtex4-1}
\pdfoutput=1
\usepackage[utf8]{inputenc}
\usepackage{hyperref}
\usepackage{graphicx}
\usepackage{amsmath,amssymb,graphicx}
\usepackage{slashed}

\begin{document}

\title{
FIMPzilla dark matter and conformal sectors
}

\author{Ameen Ismail}
\email{ai279@cornell.edu}
\affiliation{Department of Physics, LEPP, Cornell University, Ithaca, NY 14853, USA}

\begin{abstract}
    We point out a dark matter candidate which arises in a minimal extension of solutions to the hierarchy problem based on compositeness. In such models, some or all of the Standard Model fields are composites of a conformal field theory (CFT) which confines near the electroweak scale. We posit an elementary scalar field, whose mass is expected to lie near the cutoff of the CFT, and whose couplings to the Standard Model are suppressed by the cutoff. Hence it can naturally be ultraheavy and feebly coupled. This scalar can constitute all of the dark matter for masses between $10^{10}$~GeV and $10^{18}$~GeV, with the relic abundance produced by the freeze-in mechanism via a coupling to the CFT. The principal experimental constraints come from bounds on the tensor-to-scalar ratio. We speculate about future detection prospects.
\end{abstract}
\maketitle

\section{Introduction}
The preponderance of evidence for dark matter (DM) provides strong experimental motivation for new physics beyond the Standard Model (SM). Despite this the microscopic nature of DM remains unknown, with a range of about 80 orders of magnitude allowed for the mass of its dominant component~\cite{Hu:2000ke,Villanueva-Domingo:2021spv}. Conversely, one of the strongest theoretical motivations for new physics is the Higgs hierarchy or naturalness problem. Within the SM the Higgs mass is a relevant operator not protected by any symmetries, and thus it is unstable against radiative corrections.

For several decades there has been speculation of a connection between DM and the hierarchy problem. Most solutions to the hierarchy problem involve new particles appearing near the TeV scale, and if they are stable they can constitute the dark matter. These ``weakly interacting massive particles'' (WIMPs) typically have weak-scale couplings to the SM particles. Such couplings are of the right size to generate the observed DM relic abundance via thermal freeze-out, a numerical coincidence dubbed the ``WIMP miracle''.
The WIMP paradigm has been extensively studied in the literature, manifesting as various avatars in popular solutions to the hierarchy problem~\cite{Jungman:1995df,Servant:2002aq,Cheng:2002ej,Agashe:2007jb,Birkedal:2006fz}. Examples include the lightest Kaluza--Klein particle in extra-dimensional settings, stabilized by KK-parity~\cite{Servant:2002aq,Cheng:2002ej}, and the lightest neutralino in the minimal supersymmetric standard model, stabilized by $R$-parity~\cite{Jungman:1995df}. (Imposing $R$-parity is also motivated by phenomenological bounds on proton decay, which perhaps renders the  WIMP miracle more miraculous.) But ever-tightening direct detection bounds have now ruled out many simple WIMP models~\cite{Arcadi:2017kky,LZ:2022ufs}. This has encouraged intense exploration of other DM frameworks beyond WIMPs, with varying degrees of theoretical motivation. For example, ultralight DM has seen a flurry of activity in recent years (for reviews see~\cite{Ferreira:2020fam,Antypas:2022asj,Chadha-Day:2021szb}), with the prototypical candidate being the QCD axion, whose existence is motivated by the strong CP problem~\cite{Peccei:1977hh,Weinberg:1977ma}.

For the purposes of this paper we will be concerned with ultraheavy particle DM, roughly defined as DM with a mass above the WIMP unitarity bound of $\sim 100$~TeV but below the Planck scale. This regime has received relatively little attention, perhaps due to a lack of model-building efforts and the experimental challenges of detection. For a recent review including existing models and efforts for detection, see~\cite{Carney:2022gse}.

In this work we draw attention to an ultraheavy DM candidate that arises in the context of solutions to the hierarchy problem based on a conformal sector. This popular paradigm (reviewed in~\cite{Bellazzini:2014yua,Panico:2015jxa}) posits that the Higgs, as well as possibly the SM gauge bosons and fermions, are composites of a conformal field theory (CFT) that confines around the TeV scale. Scale invariance allows for a large, radiatively stable hierarchy between the ultraviolet (UV) cutoff of the CFT and the infrared (IR) scale at which it confines, protecting a small Higgs mass. Our key idea is that if one introduces a scalar field which is elementary (as opposed to being a composite of the CFT), its mass will naturally lie close to the UV cutoff, which may easily be ultraheavy. This scalar can constitute the DM, with the relic abundance set by the freeze-in mechanism~\cite{Hall:2009bx,Elahi:2014fsa}. The initial abundance after inflation is negligible, and the DM is generated through a coupling to the CFT, while the CFT is still in its hot, deconfined phase.

One could also describe the same scenario in the 5D dual picture through the AdS/CFT correspondence~\cite{Arkani-Hamed:2000ijo,Rattazzi:2000hs}. Composite Higgs models are dual to warped extra dimensions, involving a slice of 5D AdS capped by UV and IR branes~\cite{Randall:1999ee,Contino:2003ve} (see also~\cite{Erdmenger:2020lvq} for a recent study of 5D duals of composite models). The Higgs and possibly the other SM fields propagate on the IR brane or in the 5D bulk. The model we are suggesting corresponds to a scalar DM candidate which lives on the UV brane.

For our mechanism to work and be natural, it is essential that the Higgs is composite. Otherwise there would be a renormalizable coupling of the Higgs to the DM, which would lead to an unstable hierarchy between the electroweak scale and the ultraheavy DM mass, as well as possibly affecting the DM abundance. We will see that the fermions and gauge bosons may be elementary, depending on the details of reheating.

The DM has small scattering cross sections with SM particles, since they are highly suppressed by the UV scale or the large mass. Such weak interactions are reminiscent of models of feebly interacting massive particles (FIMPs)~\cite{Hall:2009bx,Bernal:2017kxu}. FIMPs have been previously studied in the context of warped extra dimensions, but only in the case where the DM propagates on the IR brane or in the bulk. If the SM lives on the UV brane but the DM is in the bulk, the 4D dual describes an elementary SM and composite DM, which is basically opposite in spirit to this work. This setup has been explored in the conformal freeze-in scenario~\cite{Hong:2019nwd,Hong:2022gzo,Chiu:2022bni}, as well as with non-FIMP DM in warped/conformal dark sectors~\cite{Blum:2014jca,vonHarling:2012sz,McDonald:2012nc,McDonald:2010fe,McDonald:2010iq,Brax:2019koq}. (See also continuum freeze-out, which features a WIMP-like DM state living in the bulk while the SM lies on the UV brane~\cite{Csaki:2021gfm,Csaki:2021xpy,Csaki:2022lnq,Fichet:2022ixi,Fichet:2022xol}.) Another case that has been studied in the literature is the SM and DM both localized on the IR brane, with a feeble SM-DM coupling~\cite{Bernal:2020fvw,Bernal:2020yqg,deGiorgi:2021xvm,deGiorgi:2022yha}.
One can also consider IR-localized WIMPs in a composite Higgs setting, which arise naturally as pseudo-Nambu--Goldstone bosons in models with extended global symmetry~\cite{Frigerio:2012uc,Chala:2012af,Marzocca:2014msa,Fonseca:2015gva,Kim:2016jbz,Wu:2017iji,Ballesteros:2017xeg,Balkin:2017aep,Alanne:2018xli,Balkin:2018tma,Cacciapaglia:2019ixa,Ramos:2019qqa,Cai:2020njb}.
What we are proposing here is fundamentally different. Our DM is elementary, and thus naturally ultraheavy and feebly coupled to the SM. Since it is heavier than a typical FIMP it is fitting to call it a ``FIMPzilla'', in analogy with WIMPzillas. FIMP\-zillas have been considered elsewhere in the literature; the best-known model is probably Planckian interacting DM, involving Planck-scale DM interacting just gravitationally with the SM~\cite{Garny:2015sjg,Garny:2017kha}. FIMPzillas with a mass around $10^{10}$~GeV in the context of a seesaw mechanism for neutrino masses were discussed in~\cite{Chianese:2019epo}.

In what follows we illustrate our ideas with a minimal model and then explore its phenomenology. The calculation of the relic abundance is complicated by the fact that the DM freezes in before the conformal phase transition occurs, when the theory is described by a hot CFT coupled to the DM. To compute the production rate we leverage techniques previously used to study scale-invariant sectors in the context of unparticle physics~\cite{Georgi:2007ek}. Our simple model successfully reproduces the observed DM relic abundance while being consistent with experimental bounds for masses up to one or two orders of magnitude below the Planck scale. The principal experimental constraint on the model arises from limits on the tensor-to-scalar ratio. We speculate about other approaches for detection, including the cosmological collider and direct gravitational detection.

\section{Model}
We suppose that the SM degrees of freedom are composites of a conformal sector which confines not too far above the electroweak scale. The dark matter is an elementary scalar, SM singlet $\chi$, which is naturally expected to lie near the cutoff scale of the CFT $\Lambda$. We require a mild hierarchy between the mass of $\chi$ and the cutoff, $m \lesssim \Lambda$, so that we can reliably perform calculations in the effective theory, without the need for a UV completion. We assume that $\chi$ is odd under a discrete $Z_2$ symmetry so that it is stable. Lastly, since we eventually will set the relic abundance of $\chi$ through the freeze-in mechanism, we assume that the initial abundance of $\chi$ is negligible. This is possible if, for example, inflationary dynamics only reheat the conformal sector.

Ultimately, only the Higgs needs to be composite to stabilize a large hierarchy. In realistic models addressing the hierarchy problem (e.g. the minimal composite Higgs~\cite{Agashe:2004rs}), one typically takes some of the SM fields to be composite while others, particularly the light fermions, are elementary. It is quite feasible to construct a composite Higgs model in which all of the SM fermions and gauge bosons are elementary~\cite{Agashe:2003zs}. If any elementary SM states are reheated after inflation, they could contribute to and even dominate DM production. This is not the situation we are interested in, so for now we work under the assumption that \textit{only the conformal sector} is reheated. Then taking some of the SM particles to be elementary has no substantial effect on DM production. Later we will relax this assumption and consider the conditions for production from the CFT sector to dominate.

In the confined phase of the CFT, the holographic dual of our setup is essentially a Randall--Sundrum model~\cite{Randall:1999ee,Agashe:2003zs} --- a slice of 5D AdS space where the composite and elementary fields are respectively localized toward the IR and UV branes --- plus a scalar $\chi$ which propagates on the UV brane.
 The cutoff of the CFT is identified with the location of the UV brane, as well as the inverse AdS curvature of the bulk. At high temperatures, the CFT is in its deconfined phase, and the dual theory is instead described by AdS-Schwarzschild space with a UV brane (where $\chi$ is still localized). The usual AdS/CFT relations allow us to write the number of colors of the CFT, $N$, in terms of the unreduced Planck scale $M_P$ and the cutoff~\cite{vonHarling:2017yew}:
\begin{equation}\label{eq:holographic}
    N = \sqrt{2\pi} \frac{M_P}{\Lambda} .
\end{equation}
The holographic theory is only under theoretical control in the large $N$ limit, $N \gg 1$. Note also that the 5D Planck scale $M_5$ is related to the 4D Planck scale and the cutoff as $M_5^3 = M_P^2 \Lambda / 8 \pi$. We will not make any further use of the holographic dual than this, but it is still a useful picture to keep in mind.

The relic abundance of $\chi$ is generated via the freeze-in mechanism in the early universe, when the CFT is in its hot, deconfined phase. We introduce a coupling of the DM to the CFT sector through an effective interaction of the form
\begin{equation}\label{eq:coupling}
    \mathcal{L}_{\rm int} = \frac{\kappa}{2} \frac{\chi^2 \mathcal{O}_{\rm CFT}}{\Lambda^{d - 2}}
\end{equation}
where $\kappa$ is a dimensionless coupling and $\mathcal{O}_{\rm CFT}$ is a CFT operator of dimension $d$. In the 5D picture this corresponds to a UV brane-localized coupling. DM can be produced through the process ${\rm CFT~stuff} \rightarrow \chi\chi$.

Note that the operator dimension $d$ need not be an integer. We do require $d > 1$ from CFT unitarity. Also when the operator is relevant, $d < 4$, $\mathcal{O}_{\rm CFT}$ needs to be forbidden from appearing in the Lagrangian by itself --- otherwise we would have a large explicit and relevant symmetry-breaking term, causing the CFT description to break down not far below the UV scale. The details of how this is achieved are not important to our mechanism, but one way would be to charge $\mathcal{O}_{\rm CFT}$ under a discrete symmetry.

In the next section we will perform a detailed calculation of the relic abundance resulting from the coupling in Eq.~\eqref{eq:coupling}. Technically we have an IR freeze-in scenario for $d \leq 2$ and UV freeze-in for $d > 2$. However it turns out that to get the right relic abundance the reheating temperature must be less than the DM mass. The hallmark of IR freeze-in is DM production dominantly occurring at $T \sim m$, but for our case the temperature is never this large even at early times. Instead the DM production rate is always suppressed by Boltzmann factors $e^{-m/T}$, so production dominantly occurs at early times, which is characteristic of UV freeze-in. Hence, the situation resembles UV freeze-in even when $d \leq 2$.

The DM mass is naturally expected to lie near the UV cutoff scale (or perhaps a loop factor below). Moreover, the UV cutoff can in principle be as large as desired, so long as the energy densities involved are sub-Planckian ($m^2 \Lambda^2 \lesssim M_P^4$)\footnote{A slightly stronger but safer condition is to simply require $\Lambda < M_P$. Super-Planckian values of $\Lambda$ raise difficulties in UV-completing to a theory of quantum gravity, but these issues can be avoided if the large size of $\Lambda$ is generated by, for example, clockwork~\cite{Choi:2015fiu,Kaplan:2015fuy,Giudice:2016yja} or monodromy~\cite{Silverstein:2008sg}. In any case this point has no qualitative bearing on our model.}, and thus the DM can easily be ultraheavy. In fact, the benchmark points we will study turn out to only be experimentally viable for masses larger than about $10^{10}$~GeV. The DM interacts very feebly with SM particles because the cross sections are suppressed by its mass (for $d > 2$ the DM-SM cross section is further suppressed by the cutoff scale). Because of the FIMP-like interactions but large mass, it is appropriate to refer to $\chi$ as a FIMPzilla.

Furthermore, the weak coupling and large mass make it infeasible to detect this DM through direct or indirect detection. Instead the principal experimental bounds arise from constraints on the CMB. In particular, we will see that the requirement of matching the observed DM relic abundance makes sharp predictions for the reheating temperature, allowing us to constrain the model through the tensor-to-scalar ratio in the CMB power spectrum. In this regard the phenomenology of our model is most similar to the Planckian interacting dark matter scenario~\cite{Garny:2015sjg}.

\section{Relic abundance}

We now proceed to calculate the DM relic abundance in this model. For completeness, let us first write the Lagrangian for $\chi$:
\begin{equation}
    \mathcal{L}_\chi = \frac{1}{2} \left( \partial \chi \right)^2 - \frac{1}{2} m^2 \chi^2 + \frac{\kappa}{2} \frac{\chi^2 \mathcal{O}_{\rm CFT}}{\Lambda^{d - 2}} .
\end{equation}
As usual for freeze-in DM, we assume the initial abundance of $\chi$ is negligible, so we can ignore annihilation of $\chi$ into CFT stuff. The Boltzmann equation for $\chi$ is then
\begin{equation}\label{eq:boltzmann}
    \dot{n}_\chi + 3 H n_\chi = n^{\rm eq}_{\rm CFT} \langle \sigma v ({\rm CFT} \rightarrow \chi\chi) \rangle
\end{equation}
where $n$ denotes a number density, $n^{\rm eq}$ an equilibrium number density, $\langle \sigma v \rangle$ a thermally averaged cross section, and $H$ is the Hubble parameter. The right-hand side of this equation should be understood as a schematic representation of processes which produce DM pairs out of the hot CFT. It should not be taken literally --- it is unclear how to even define a scattering process with CFT stuff in the initial state, due to the lack of asymptotic states in a CFT.

Nevertheless, one can make sense of the production rate by relating it to the inverse process $\chi\chi \rightarrow {\rm CFT}$ using the principle of detailed balance:
\begin{equation}
    n^{\rm eq}_{\rm CFT} \langle \sigma v({\rm CFT} \rightarrow \chi\chi) \rangle = \left( n_\chi^{\rm eq}\right )^2 \langle \sigma v(\chi\chi \rightarrow {\rm CFT}) \rangle \equiv \gamma .
\end{equation}
The rate $\gamma$ can be computed using unparticle methods for the CFT phase space~\cite{Georgi:2007ek}; we provide the calculational details in the appendix. One could also calculate the rate using the techniques in~\cite{Redi:2021ipn}. (Presumably, we could perform this calculation in the 5D picture too, by considering an AdS-Schwarzschild metric with a UV brane on which $\chi$ propagates.) One finds
\begin{equation}\label{eq:CFTrate}\begin{split}
    \gamma &= \kappa^2 A_d \frac{T}{32\pi^4} (2m)^3 \left( \frac{2m}{\Lambda} \right)^{2d-4} \\
    &\times \int_1^\infty du u^{2d-4} \sqrt{u^2-1} K_1(2mu/T)
\end{split}\end{equation}
where $A_d = 16 \pi^{5/2} (2\pi)^{-2d} \Gamma(d+1/2)/ (\Gamma(d-1) \Gamma(2d))$ is the usual unparticle phase space normalization, and $K_1$ is a modified Bessel function of the second kind.

It is convenient to write the Boltzmann equation (Eq.~\eqref{eq:boltzmann}) in terms of the abundance $Y = n_\chi/s$ (where $s$ is the entropy density) and $x = m/T$, leading to
\begin{equation}
    \frac{dY}{dx} = \frac{\gamma}{H s x} .
\end{equation}
We then substitute in Eq.~\ref{eq:CFTrate} for $\gamma$, use $H = 1.66 \sqrt{g_*} T^2/M_P$ and $s = 2\pi^2 g_* T^3/45$, and integrate the Boltzmann equation to obtain the DM abundance today:
\begin{equation}\begin{split}
    Y_\infty &= \kappa^2 \frac{45 A_d}{(1.66) 8 \pi^6 g_*^{3/2}} \frac{M_P}{m} \left( \frac{2m}{\Lambda} \right)^{2d-4} \\
    &\times \int_{x_R}^\infty dx \int_1^\infty du u^{2d-4} \sqrt{u^2-1} x^3 K_1(2xu) .
\end{split}\end{equation}
Here $x_R$ is the value of $x$ at the reheating temperature $T_R$: $x_R = m/T_R$. Also, when the CFT is in the hot phase the effective number of relativistic degrees of freedom is controlled by the number of colors in the CFT, $g_* \sim N^2 = 2\pi M_P^2/\Lambda^2$.

The relic abundance is sensitive to the reheating temperature, which is typical of UV freeze-in. Assuming efficient reheating, the reheating temperature is related to the Hubble parameter at the end of inflation $H_I$ as
\begin{equation}
    H_I = \sqrt{\frac{4\pi^3 g_*}{45}} \frac{T_R^2}{M_P} .
\end{equation}
This is important because $H_I$ is constrained by the CMB bound on the tensor-to-scalar ratio $r$.

\section{Phenomenology}

\begin{figure}
    \includegraphics[width=\columnwidth]{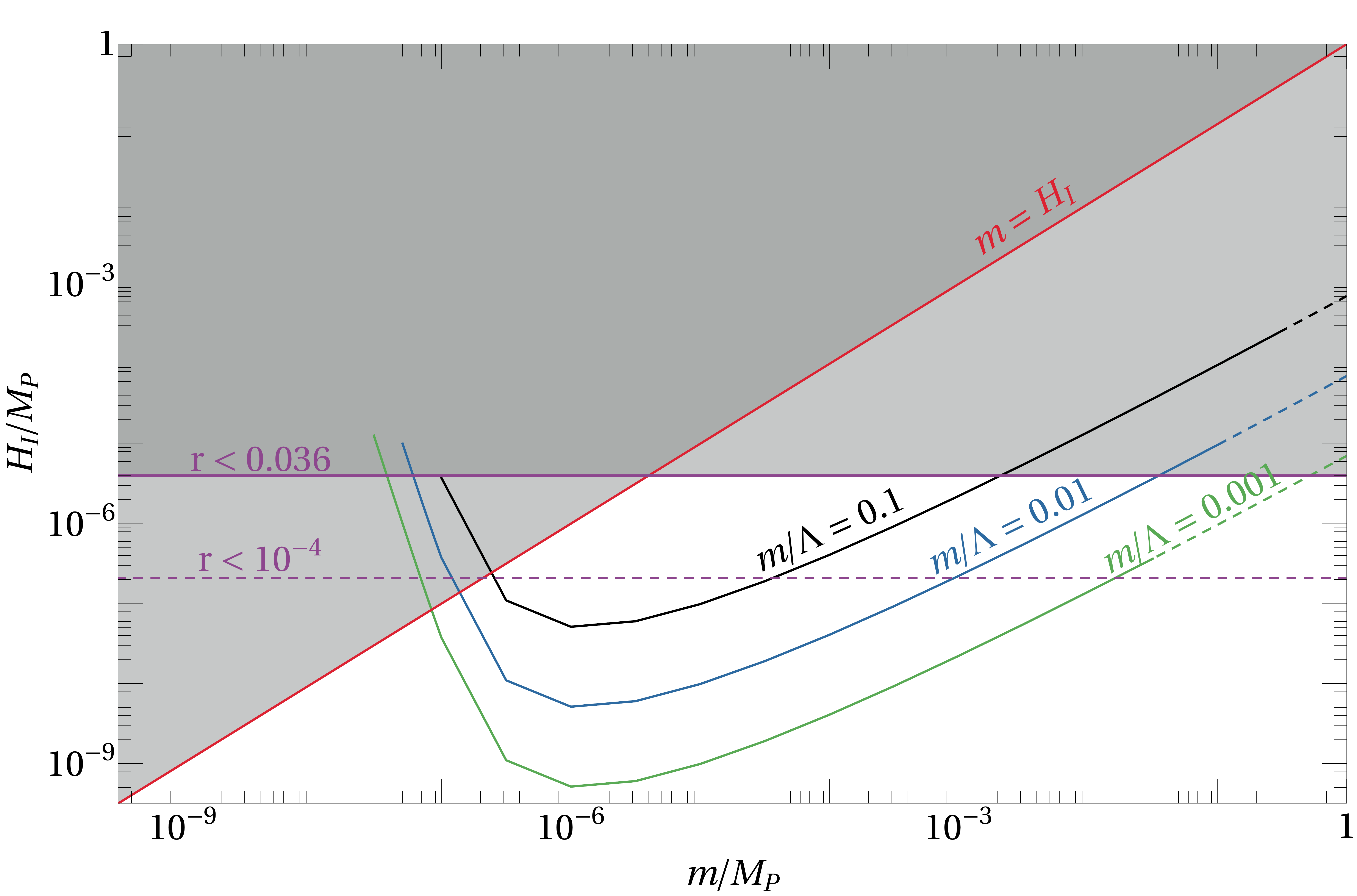}
    \caption{The value of $H_I$ which yields the observed DM relic abundance~\cite{Planck:2018vyg,ParticleDataGroup:2022pth} as a function of the DM mass $m$. We fix $d = 3.5$ and $\kappa = 1$. We show results for $m/\Lambda = 0.1$ \textit{(black line)}, $m/\Lambda = 0.01$ \textit{(blue line)}, and $m/\Lambda = 0.001$ \textit{(green line)}. The dashing of these lines indicates where the energy density is super-Planckian. We show bounds from the nonobservation of isocurvature perturbations, which excludes $m < H_I$ \textit{(red line)}~\cite{Garny:2015sjg,Planck:2018jri}; from the CMB upper limit on the tensor-to-scalar ratio,  $r < 0.036$ \textit{(purple line)}~\cite{BICEP:2021xfz,ParticleDataGroup:2022pth}; and a projection for a future bound $r < 10^{-4}$ \textit{(dashed purple line)}.}
    \label{fig:varyingmass}
\end{figure}

In Fig.~\ref{fig:varyingmass} we show curves in the $(m, H_I)$ plane which yield the observed DM relic abundance~\cite{Planck:2018vyg,ParticleDataGroup:2022pth}. We fix $d = 3.5$ and $\kappa = 1$ and vary $m/\Lambda = 0.1,0.01,0.001$ over the three curves. We indicate the maximally allowed value of $H_I$ from the CMB bound on the tensor-to-scalar ratio $r < 0.036$~\cite{BICEP:2021xfz,ParticleDataGroup:2022pth}. We also include a bound $m > H_I$, which safely avoids DM isocurvature perturbations~\cite{Garny:2015sjg,Planck:2018jri}.

Note that the DM mass cannot be taken arbitrarily close to the Planck scale. Quantum gravity effects become important as the energy density of $\chi$, which is of order $m^2 \Lambda^2$, becomes comparable to $M_P^4$. For this reason we have dashed the curves in Fig.~\ref{fig:varyingmass} for $m/M_P > \sqrt{m/\Lambda}$, where our calculations are not reliable. Also, there is a minimum mass below which the relic abundance is too small no matter how large the reheating temperature is, which is why the curves do not extend below $m \sim 10^{-7} M_P$. We further caution that we have neglected direct gravitational production of $\chi$, which may be important when $m$ is very close to $H_I$~\cite{Garny:2015sjg}.

Together, the benchmark points in Fig.~\ref{fig:varyingmass} provide experimentally viable DM candidates for masses ranging from $10^{-7} M_P \sim 10^{12}$~GeV to $0.1 M_P \sim 10^{18}$~GeV. As previously stated, it is difficult to search for the DM with direct or indirect detection because it is ultraheavy and couples feebly to the SM. However, the model is in principle testable through the bound on the tensor-to-scalar ratio. As the bound on $r$ becomes tighter, one needs to fine-tune $m/\Lambda$ to smaller values to evade the bound. In Fig.~\ref{fig:varyingmass} we show a projection for a bound $r < 10^{-4}$, which would probe a substantial region of parameter space. (The specific choice of $10^{-4}$ was quoted as a ``futuristic bound'' in~\cite{Garny:2015sjg}; for comparison, the upcoming CMB-S4 experiment is projected to be sensitive to $r$ down to about $10^{-3}$ to $10^{-4}$~\cite{CMB-S4:2016ple}.)

\begin{figure}
    \includegraphics[width=\columnwidth]{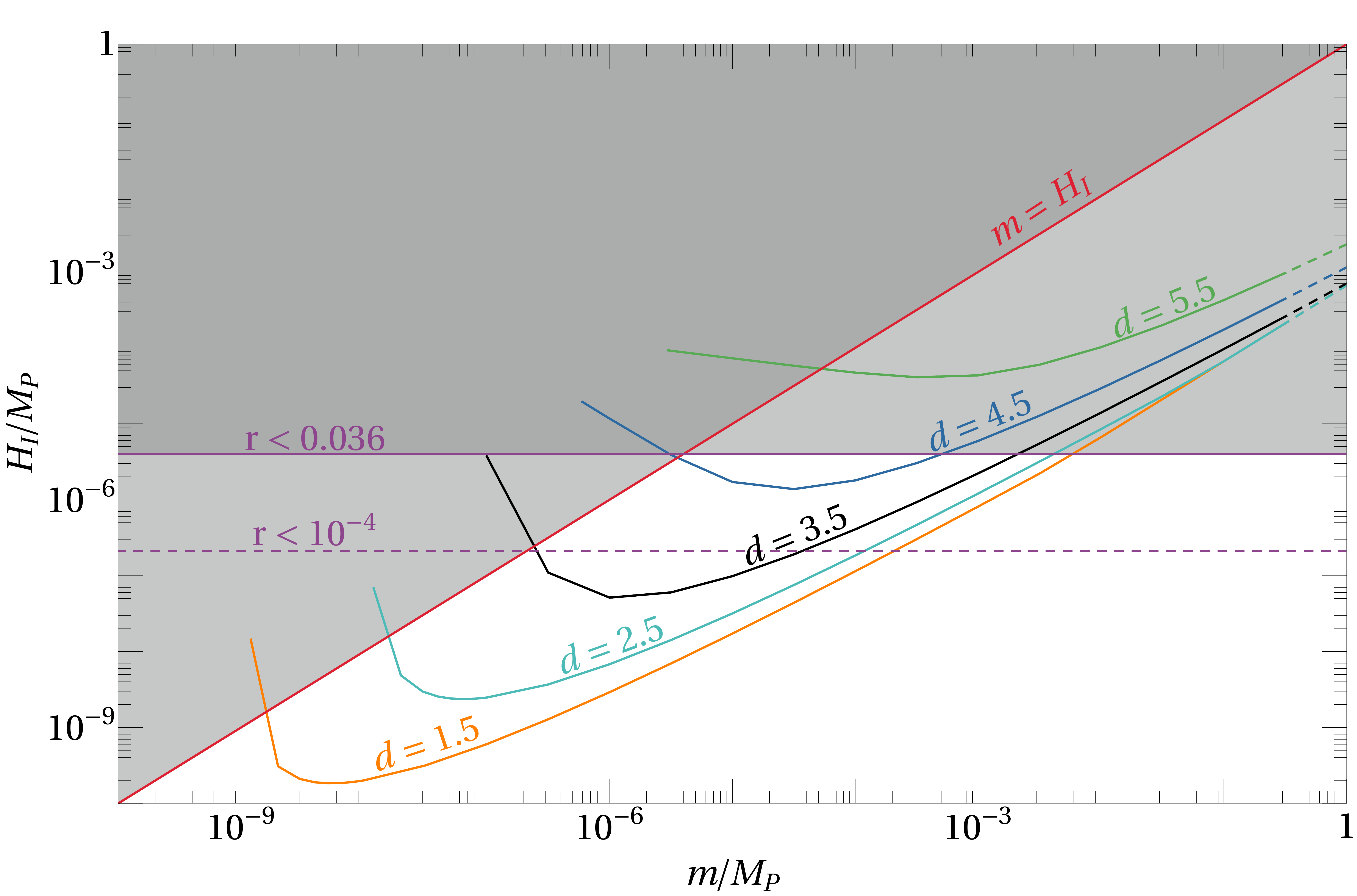}
    \caption{Same as Fig.~\ref{fig:varyingmass}, but fixing $m/\Lambda = 0.1$ and $\kappa = 1$ while choosing different values of $d$: $d = 1.5$ \textit{(orange line)}, $d = 2.5$ \textit{(turquoise line)}, $d = 3.5$ \textit{(black line)}, $d = 4.5$ \textit{(blue line)}, and $d = 5.5$ \textit{(green line)}.}
    \label{fig:varyingdim}
\end{figure}

Fig.~\ref{fig:varyingdim} illustrates the effect of changing the operator dimension, fixing $m/\Lambda = 0.1$ and $\kappa = 1$ while varying $d$ from $1.5$ to $5.5$. The choice of half-integer values is merely to emphasize that $d$ does not need to be an integer. The bounds are the same as in Fig.~\ref{fig:varyingmass}. As $d$ increases, the DM production rate is suppressed by a larger power of the cutoff $\Lambda$, and so obtaining the correct relic abundance requires a larger reheating temperature and thus a larger $H_I$.

For $d \gtrsim 5$ it is difficult to satisfy the existing bound on $r$ while generating the right relic abundance. Since CFT unitarity implies $d > 1$, there is a narrow permitted window for the operator dimension $1 < d \lesssim 5$ (the exact upper bound depends weakly on the value of $m/\Lambda$). This provides a complementary view on how constraining the tensor-to-scalar ratio probes our model: as the maximum allowed $r$ becomes smaller, the viable window for the operator dimension $d$ shrinks.

We can also see from Fig.~\ref{fig:varyingdim} that lowering the dimension allows for a smaller DM mass, although still firmly in the ultraheavy regime; for $d = 1.5$ (and $m/\Lambda = 0.1$), $m$ can be as low as $\sim 10^{-9} M_P$. One might worry about direct detection bounds becoming important for $d < 2$, as the coupling between $\chi$ and the CFT becomes relevant. Recall the DM couples to the CFT like $\Lambda^{2-d} \chi^2 \mathcal{O}_{\rm CFT}$, enhanced by powers of $\Lambda$ when $d < 2$. To obtain the resulting interaction with a composite fermion $\psi$, we match the CFT operator onto the low-energy theory as $\mathcal{O}_{\rm CFT} \rightarrow \sqrt{N}/(4 \pi \Lambda_{\rm IR}^{3-d}) \psi^\dagger \psi$, where $\Lambda_{\rm IR}$ is the scale at which the CFT confines. This scaling is suggested by dimensional analysis, Lorentz invariance, and large-$N$ arguments~\cite{Hong:2022gzo}. We can then calculate cross sections for direct detection in the normal way. Taking $\Lambda_{\rm IR} \sim$~TeV and using the holographic relation Eq.~\eqref{eq:holographic}, we obtain a DM-nucleon cross section of
\begin{equation}
    \sigma \simeq 10^{-75} {\rm \:cm^2\:} \frac{f_N^2 m}{\Lambda} \left( \frac{\Lambda}{1{\rm \:TeV}} \right)^{2-d}  \left( \frac{M_P}{m} \right)^3
\end{equation}
where $f_N$ is an $\mathcal{O}(1)$ nuclear form factor. Even for the smallest masses in Fig.~\ref{fig:varyingdim} (that is, $m \sim 10^{-9} M_P$ and $d = 1.5$), the cross section is of order $10^{-45}$~${\rm cm}^2$, well below direct detection bounds.

Lastly, we speculate about further opportunities for detection. For masses near the inflationary Hubble scale, $m \sim H_I$, one could potentially see imprints of $\chi$ in cosmological collider observables~\cite{Arkani-Hamed:2015bza}. A more involved analysis than what we have done in this work is required to understand the details of this. In particular one would need to take into account gravitational production of DM when computing the relic abundance, since that is important in the very regime $m \sim H_I$ relevant for the cosmological collider\footnote{This is really a trivial statement. Particles with masses near the Hubble scale have interesting effects on cosmological correlators \textit{because} they can be produced during inflation.}. Another compelling avenue for detection is to directly probe the gravitational coupling of $\chi$ to ordinary matter. This is the ultimate goal pursued by the Windchime project: gravitational detection of Planck-scale DM~\cite{Carney:2019pza,Windchime:2022whs}. Our model provides another physics case for Windchime in the form of a well-motivated, ultraheavy particle DM candidate which is difficult to detect through other means.

\section{Elementary production}
As we emphasized earlier, we have heretofore assumed that any elementary SM particles are not reheated. We now relax this assumption, which allows for additional freeze-in DM production from annihilation of SM fermions or gauge bosons to $\chi\chi$. We want to consider under what circumstances these processes provide the leading contribution to DM production.

An elementary SM fermion $\psi$ or gauge boson $A_\mu$ can couple to $\chi$ through a dimension-six operator:
\begin{equation}\label{eq:fermioncoupling}
    \mathcal{L}_{\rm fermion} = \frac{\kappa_f}{2\Lambda^2} \overline{\psi} \slashed{D} \psi \chi^2, \quad \mathcal{L}_{\rm vector} = \frac{\kappa_v}{4\Lambda^2} F_{\mu\nu} F^{\mu\nu} \chi^2 .
\end{equation}
If $\chi^2$ can couple to a CFT operator with dimension $1 < d < 4$ (see Eq.~\eqref{eq:coupling}), then $\chi\chi$ production from fermions and gauge bosons is suppressed by more powers of $\Lambda$ than CFT production. Hence we expect CFT production to dominantly produce the DM relic abundance. On the other hand, if $d > 4$, then production from gauge bosons, $A A \rightarrow \chi\chi$, dominates. (We argue in the appendix that production from fermions is subdominant.)

For completeness, in the appendix we study the latter scenario in the extreme case where all SM particles are elementary and reheated at the end of inflation. This is a standard UV freeze-in calculation, in contrast to the main focus of this paper, CFT-dominated production, where we had to use unparticle tricks. In this case $\chi$ can still be a viable DM candidate for masses between $10^{-8} M_P$ and $10^{-2} M_P$.

We reiterate that freeze-in production from elementary SM particles is only pertinent when those particles are reheated at the end of inflation. If only the CFT is reheated, the elementary fields do not impact DM production. The central conclusion of this discussion is that even when some elementary fields are reheated, CFT production may dominate anyway if $d < 4$, in which case the results of the previous section still apply.

\section{Discussion}
We have demonstrated here that solutions to the hierarchy problem based on a conformal sector easily accommodate an ultraheavy and feebly-interacting DM candidate in the form of an elementary scalar.
Focusing on a minimal realization of this scenario, we studied the case where freeze-in production of the DM occurs dominantly through a coupling to the CFT. This led to a viable DM candidate for ultraheavy masses upwards of $10^{-9} M_P \sim 10^{10}$~GeV.

The principal experimental constraints arose from the CMB bound on the tensor-to-scalar ratio $r$, which bounds the inflationary Hubble scale. In the future the parameter space will be further constrained by improved bounds on $r$ from experiments like CMB-S4. It is possible that the cosmological collider could probe masses not too far above the inflationary Hubble scale, although we leave a detailed study of this for future work. Larger masses close to the Planck scale could provide an additional physics case for proposals for direct gravitational detection.

The importance of the conformal symmetry to our model should not be understated. Otherwise, one might simply posit an ultraheavy Dirac fermion as a DM candidate; it would only have nonrenormalizable couplings to the SM and thus one could obtain the right relic abundance through UV freeze-in. But this misses the point: there is a large hierarchy between the UV scale at which the ultraheavy DM lies and the IR scale at which the SM particles lie. This hierarchy is fine-tuned in the absence of any symmetries protecting it. The essential role of the conformal sector is to dynamically generate a large and stable UV/IR hierarchy.

Throughout this work we have neglected any discussion of the conformal phase transition, and in fact we have tacitly assumed that the phase transition occurs promptly, without supercooling. A large amount of supercooling would dilute the DM abundance, potentially posing problems for our production mechanism\footnote{Possibly, one could evade this issue by overproducing DM in the early universe, such that after being diluted in a period of supercooling, one is left with the right relic abundance. It would be interesting to study how much supercooling can be accommodated in this way.}. Whether a prompt phase transition can be achieved is dependent on the stabilization mechanism; the prototypical Goldberger--Wise mechanism often leads to a supercooled transition~\cite{Goldberger:1999uk,Creminelli:2001th,Randall:2006py,Konstandin:2011dr,Baratella:2018pxi}. Nevertheless, alternative approaches to stabilization which avoid supercooling have been proposed~\cite{Hassanain:2007js,Bunk:2017fic,Dillon:2017ctw,Megias:2018sxv,Agashe:2019lhy,Csaki:2023pwy,Eroncel:2023uqf}. Another option is that the CFT is always in the confined phase, as suggested in~\cite{Agrawal:2021alq}. This would alter the calculation of the relic abundance, however, since we assumed freeze-in occurs when the CFT is in the deconfined phase.

We have also remained largely agnostic about the details of inflation. Freeze-in requires that the initial abundance of the DM is negligible. The obvious way to achieve this is to assume that the inflationary dynamics cause only the conformal sector to be reheated. There is a rich literature on inflation in warped throats, mainly in the context of string theory (e.g.~\cite{Kachru:2003sx,Bassett:2005xm,Douglas:2006es,McAllister:2008hb,Baumann:2014nda}). In light of this, it would be interesting to study the interplay of inflationary model-building with our DM model --- perhaps this could lead to additional phenomenological signatures. Furthermore, one could explore the scenario in which some of the SM fermions and/or gauge bosons are elementary, but still only the conformal sector is reheated. Would this have any consequences for the cosmological history of these models? We leave all these exciting questions for future projects.

\begin{acknowledgments}
    We are very grateful to Maxim Perelstein for several fruitful discussions, useful comments, and encouragement. We also thank Csaba Cs\'aki, Sylvain Fichet, Jay Hubisz, and Seung J. Lee for helpful comments. We are supported in part by the NSF grant PHY-2014071 and in part by NSERC, funding reference number 557763.
\end{acknowledgments}

\appendix

\section{DM production calculation}\label{app:A}

Here we provide some further details of the calculation of Eq.~\eqref{eq:CFTrate}. We first calculate the cross section $\sigma$ for the processs $\chi\chi \rightarrow {\rm CFT}$. The CFT phase space factor is given in~\cite{Georgi:2007ek}, and we find
\begin{equation}\begin{split}
    \sigma &= \frac{A_d}{4\sqrt{ \left( p_1 \cdot p_2 \right)^2 - m^4 }}  \\
    &\times \int \frac{d^4 p}{(2\pi)^4} \lvert \mathcal{M} \rvert^2 (2\pi)^4 \delta^4 \left(p_1 + p_2 - p \right) \theta \left(p^0 \right) \theta \left(p^2 \right) \left( p^2 \right)^{d-2} ,
\end{split}\end{equation}
where $p_1$ and $p_2$ are the momenta of the initial $\chi$ particles, and $A_d = 16 \pi^{5/2} (2\pi)^{-2d} \Gamma(d+1/2)/ (\Gamma(d-1) \Gamma(2d))$. From Eq.~\eqref{eq:coupling} the matrix element is just $\lvert \mathcal{M} \rvert = \kappa/\Lambda^{d-2}$. The phase space integration is trivial and we are left with
\begin{equation}\label{eq:crosssection}
    \sigma = \kappa^2 A_d \left( \frac{s}{\Lambda^2} \right)^{d-2} \frac{1}{2\sqrt{s \left( s - 4m^2 \right)}}
\end{equation}
where $s = (p_1 + p_2)^2$.

We can now use the standard formula for the thermally averaged cross section~\cite{Gondolo:1990dk}:
\begin{equation}
    \left( n_\chi^{\rm eq} \right)^2 \langle \sigma v  \rangle = \frac{T}{32\pi^4} \int_{4m^2}^{\infty} ds \sigma (s-4m^2) \sqrt{s} K_1(\sqrt{s}/T) .
\end{equation}
Upon plugging in Eq.~\eqref{eq:crosssection} for the cross section (and changing integration variables to $u = \sqrt{s}/2m$), one readily obtains Eq.~\eqref{eq:CFTrate}, as desired.

\section{Freeze-in from elementary particles}\label{app:B}

Here we consider the case where production is dominated by elementary SM particles, rather than CFT production. The cross section for $AA \rightarrow \psi \psi$, where $A_\mu$ is a massless gauge boson which couples to $\chi$ as in Eq.~\eqref{eq:fermioncoupling}, is
\begin{equation}
    \sigma = \frac{\kappa_v^2}{8\pi} \left( \frac{s}{\Lambda^2} \right)^2 \frac{1}{2\sqrt{s \left( s - 4m^2 \right)}} .
\end{equation}
We remark that this is identical to the CFT production cross section, Eq.~\eqref{eq:crosssection}, with $d=4$ and the replacements $\kappa \rightarrow \kappa_v$, $A_d \rightarrow 1/8\pi$. Thus the production rate $\left( n_\chi^{\rm eq} \right)^2 \langle \sigma v \rangle$ is simply given by Eq.~\eqref{eq:CFTrate} with those replacements. 

Using the equation of motion in Eq.~\eqref{eq:fermioncoupling}, or by direct calculation, one can show that the corresponding production rate for a fermion $\psi$ must scale as $s m_\psi^2 / \Lambda^4$, so it vanishes in the massless limit. Therefore, the contribution of an elementary fermion to the DM relic abundance is suppressed with respect to the contribution from an elementary gauge boson.

\begin{figure}
    \includegraphics[width=\columnwidth]{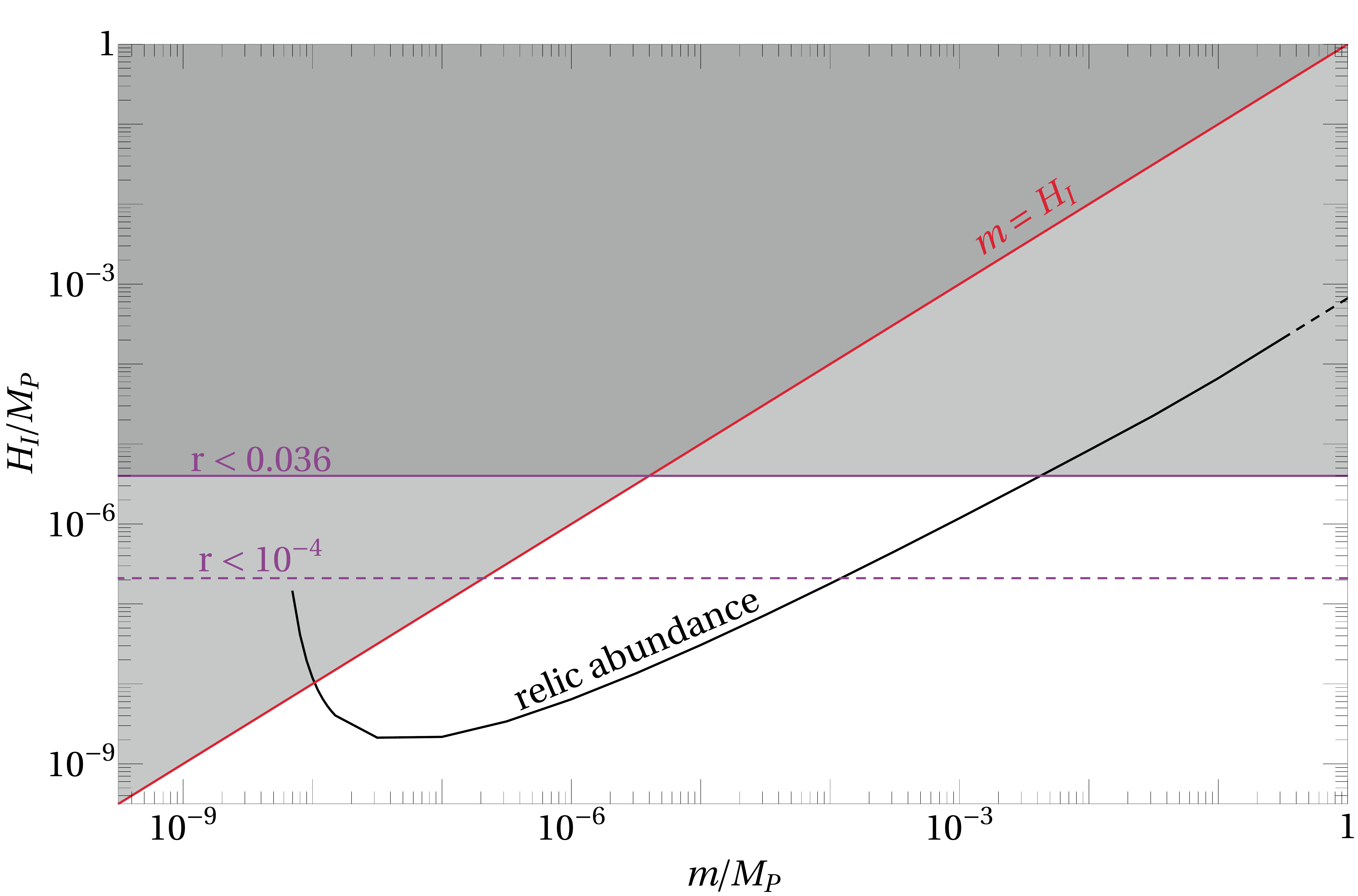}
    \caption{Values of $H_I$ and $m$ which yield the observed DM relic abundance~\cite{Planck:2018vyg,ParticleDataGroup:2022pth} in the case where freeze-in production from elementary gauge bosons dominates over CFT production \textit{(black line)}. We assume all of the SM gauge bosons are elementary, are reheated at the end of inflation, and couple universally to the DM with $\kappa_v = 1$. We fix $m/\Lambda = 0.1$. The bounds are the same as in Fig.~\ref{fig:varyingmass}.}
    \label{fig:elementary}
\end{figure}

Let us study this scenario in the case where all of the SM gauge bosons are elementary, and suppose that they all couple to $\chi$ with $\kappa_v = 1$. The production rate we just computed must be multiplied by $12$ to account for the number of gauge boson species in the SM. In Fig.~\ref{fig:elementary} we show the corresponding curve yielding the observed DM abundance, fixing $m/\Lambda = 0.1$. The bounds are the same as in Fig.~\ref{fig:varyingmass}.

From Fig.~\ref{fig:elementary}, we see this scenario reproduces the correct relic abundance while being consistent with all experimental bounds for masses between $10^{-8} M_P$ and $10^{-2} M_P$. The prospects for detection are similar to those discussed in the main text for the case where CFT production dominates.

\bibliography{references}{}

\begin{thebibliography}{95}%
\makeatletter
\providecommand \@ifxundefined [1]{%
 \@ifx{#1\undefined}
}%
\providecommand \@ifnum [1]{%
 \ifnum #1\expandafter \@firstoftwo
 \else \expandafter \@secondoftwo
 \fi
}%
\providecommand \@ifx [1]{%
 \ifx #1\expandafter \@firstoftwo
 \else \expandafter \@secondoftwo
 \fi
}%
\providecommand \natexlab [1]{#1}%
\providecommand \enquote  [1]{``#1''}%
\providecommand \bibnamefont  [1]{#1}%
\providecommand \bibfnamefont [1]{#1}%
\providecommand \citenamefont [1]{#1}%
\providecommand \href@noop [0]{\@secondoftwo}%
\providecommand \href [0]{\begingroup \@sanitize@url \@href}%
\providecommand \@href[1]{\@@startlink{#1}\@@href}%
\providecommand \@@href[1]{\endgroup#1\@@endlink}%
\providecommand \@sanitize@url [0]{\catcode `\\12\catcode `\$12\catcode
  `\&12\catcode `\#12\catcode `\^12\catcode `\_12\catcode `\%12\relax}%
\providecommand \@@startlink[1]{}%
\providecommand \@@endlink[0]{}%
\providecommand \url  [0]{\begingroup\@sanitize@url \@url }%
\providecommand \@url [1]{\endgroup\@href {#1}{\urlprefix }}%
\providecommand \urlprefix  [0]{URL }%
\providecommand \Eprint [0]{\href }%
\providecommand \doibase [0]{http://dx.doi.org/}%
\providecommand \selectlanguage [0]{\@gobble}%
\providecommand \bibinfo  [0]{\@secondoftwo}%
\providecommand \bibfield  [0]{\@secondoftwo}%
\providecommand \translation [1]{[#1]}%
\providecommand \BibitemOpen [0]{}%
\providecommand \bibitemStop [0]{}%
\providecommand \bibitemNoStop [0]{.\EOS\space}%
\providecommand \EOS [0]{\spacefactor3000\relax}%
\providecommand \BibitemShut  [1]{\csname bibitem#1\endcsname}%
\let\auto@bib@innerbib\@empty
\bibitem [{\citenamefont {Hu}\ \emph {et~al.}(2000)\citenamefont {Hu},
  \citenamefont {Barkana},\ and\ \citenamefont {Gruzinov}}]{Hu:2000ke}%
  \BibitemOpen
  \bibfield  {author} {\bibinfo {author} {\bibfnamefont {Wayne}\ \bibnamefont
  {Hu}}, \bibinfo {author} {\bibfnamefont {Rennan}\ \bibnamefont {Barkana}}, \
  and\ \bibinfo {author} {\bibfnamefont {Andrei}\ \bibnamefont {Gruzinov}},\
  }\bibfield  {title} {\enquote {\bibinfo {title} {{Cold and fuzzy dark
  matter}},}\ }\href {\doibase 10.1103/PhysRevLett.85.1158} {\bibfield
  {journal} {\bibinfo  {journal} {Phys. Rev. Lett.}\ }\textbf {\bibinfo
  {volume} {85}},\ \bibinfo {pages} {1158--1161} (\bibinfo {year} {2000})},\
  \Eprint {http://arxiv.org/abs/astro-ph/0003365} {arXiv:astro-ph/0003365}
  \BibitemShut {NoStop}%
\bibitem [{\citenamefont {Villanueva-Domingo}\ \emph
  {et~al.}(2021)\citenamefont {Villanueva-Domingo}, \citenamefont {Mena},\ and\
  \citenamefont {Palomares-Ruiz}}]{Villanueva-Domingo:2021spv}%
  \BibitemOpen
  \bibfield  {author} {\bibinfo {author} {\bibfnamefont {Pablo}\ \bibnamefont
  {Villanueva-Domingo}}, \bibinfo {author} {\bibfnamefont {Olga}\ \bibnamefont
  {Mena}}, \ and\ \bibinfo {author} {\bibfnamefont {Sergio}\ \bibnamefont
  {Palomares-Ruiz}},\ }\bibfield  {title} {\enquote {\bibinfo {title} {{A brief
  review on primordial black holes as dark matter}},}\ }\href {\doibase
  10.3389/fspas.2021.681084} {\bibfield  {journal} {\bibinfo  {journal} {Front.
  Astron. Space Sci.}\ }\textbf {\bibinfo {volume} {8}},\ \bibinfo {pages} {87}
  (\bibinfo {year} {2021})},\ \Eprint {http://arxiv.org/abs/2103.12087}
  {arXiv:2103.12087 [astro-ph.CO]} \BibitemShut {NoStop}%
\bibitem [{\citenamefont {Jungman}\ \emph {et~al.}(1996)\citenamefont
  {Jungman}, \citenamefont {Kamionkowski},\ and\ \citenamefont
  {Griest}}]{Jungman:1995df}%
  \BibitemOpen
  \bibfield  {author} {\bibinfo {author} {\bibfnamefont {Gerard}\ \bibnamefont
  {Jungman}}, \bibinfo {author} {\bibfnamefont {Marc}\ \bibnamefont
  {Kamionkowski}}, \ and\ \bibinfo {author} {\bibfnamefont {Kim}\ \bibnamefont
  {Griest}},\ }\bibfield  {title} {\enquote {\bibinfo {title} {{Supersymmetric
  dark matter}},}\ }\href {\doibase 10.1016/0370-1573(95)00058-5} {\bibfield
  {journal} {\bibinfo  {journal} {Phys. Rept.}\ }\textbf {\bibinfo {volume}
  {267}},\ \bibinfo {pages} {195--373} (\bibinfo {year} {1996})},\ \Eprint
  {http://arxiv.org/abs/hep-ph/9506380} {arXiv:hep-ph/9506380} \BibitemShut
  {NoStop}%
\bibitem [{\citenamefont {Servant}\ and\ \citenamefont
  {Tait}(2003)}]{Servant:2002aq}%
  \BibitemOpen
  \bibfield  {author} {\bibinfo {author} {\bibfnamefont {Geraldine}\
  \bibnamefont {Servant}}\ and\ \bibinfo {author} {\bibfnamefont {Timothy
  M.~P.}\ \bibnamefont {Tait}},\ }\bibfield  {title} {\enquote {\bibinfo
  {title} {{Is the lightest Kaluza-Klein particle a viable dark matter
  candidate?}}}\ }\href {\doibase 10.1016/S0550-3213(02)01012-X} {\bibfield
  {journal} {\bibinfo  {journal} {Nucl. Phys. B}\ }\textbf {\bibinfo {volume}
  {650}},\ \bibinfo {pages} {391--419} (\bibinfo {year} {2003})},\ \Eprint
  {http://arxiv.org/abs/hep-ph/0206071} {arXiv:hep-ph/0206071} \BibitemShut
  {NoStop}%
\bibitem [{\citenamefont {Cheng}\ \emph {et~al.}(2002)\citenamefont {Cheng},
  \citenamefont {Feng},\ and\ \citenamefont {Matchev}}]{Cheng:2002ej}%
  \BibitemOpen
  \bibfield  {author} {\bibinfo {author} {\bibfnamefont {Hsin-Chia}\
  \bibnamefont {Cheng}}, \bibinfo {author} {\bibfnamefont {Jonathan~L.}\
  \bibnamefont {Feng}}, \ and\ \bibinfo {author} {\bibfnamefont
  {Konstantin~T.}\ \bibnamefont {Matchev}},\ }\bibfield  {title} {\enquote
  {\bibinfo {title} {{Kaluza-Klein dark matter}},}\ }\href {\doibase
  10.1103/PhysRevLett.89.211301} {\bibfield  {journal} {\bibinfo  {journal}
  {Phys. Rev. Lett.}\ }\textbf {\bibinfo {volume} {89}},\ \bibinfo {pages}
  {211301} (\bibinfo {year} {2002})},\ \Eprint
  {http://arxiv.org/abs/hep-ph/0207125} {arXiv:hep-ph/0207125} \BibitemShut
  {NoStop}%
\bibitem [{\citenamefont {Agashe}\ \emph {et~al.}(2008)\citenamefont {Agashe},
  \citenamefont {Falkowski}, \citenamefont {Low},\ and\ \citenamefont
  {Servant}}]{Agashe:2007jb}%
  \BibitemOpen
  \bibfield  {author} {\bibinfo {author} {\bibfnamefont {Kaustubh}\
  \bibnamefont {Agashe}}, \bibinfo {author} {\bibfnamefont {Adam}\ \bibnamefont
  {Falkowski}}, \bibinfo {author} {\bibfnamefont {Ian}\ \bibnamefont {Low}}, \
  and\ \bibinfo {author} {\bibfnamefont {Geraldine}\ \bibnamefont {Servant}},\
  }\bibfield  {title} {\enquote {\bibinfo {title} {{KK Parity in Warped Extra
  Dimension}},}\ }\href {\doibase 10.1088/1126-6708/2008/04/027} {\bibfield
  {journal} {\bibinfo  {journal} {JHEP}\ }\textbf {\bibinfo {volume} {04}},\
  \bibinfo {pages} {027} (\bibinfo {year} {2008})},\ \Eprint
  {http://arxiv.org/abs/0712.2455} {arXiv:0712.2455 [hep-ph]} \BibitemShut
  {NoStop}%
\bibitem [{\citenamefont {Birkedal}\ \emph {et~al.}(2006)\citenamefont
  {Birkedal}, \citenamefont {Noble}, \citenamefont {Perelstein},\ and\
  \citenamefont {Spray}}]{Birkedal:2006fz}%
  \BibitemOpen
  \bibfield  {author} {\bibinfo {author} {\bibfnamefont {Andreas}\ \bibnamefont
  {Birkedal}}, \bibinfo {author} {\bibfnamefont {Andrew}\ \bibnamefont
  {Noble}}, \bibinfo {author} {\bibfnamefont {Maxim}\ \bibnamefont
  {Perelstein}}, \ and\ \bibinfo {author} {\bibfnamefont {Andrew}\ \bibnamefont
  {Spray}},\ }\bibfield  {title} {\enquote {\bibinfo {title} {{Little Higgs
  dark matter}},}\ }\href {\doibase 10.1103/PhysRevD.74.035002} {\bibfield
  {journal} {\bibinfo  {journal} {Phys. Rev. D}\ }\textbf {\bibinfo {volume}
  {74}},\ \bibinfo {pages} {035002} (\bibinfo {year} {2006})},\ \Eprint
  {http://arxiv.org/abs/hep-ph/0603077} {arXiv:hep-ph/0603077} \BibitemShut
  {NoStop}%
\bibitem [{\citenamefont {Arcadi}\ \emph {et~al.}(2018)\citenamefont {Arcadi},
  \citenamefont {Dutra}, \citenamefont {Ghosh}, \citenamefont {Lindner},
  \citenamefont {Mambrini}, \citenamefont {Pierre}, \citenamefont {Profumo},\
  and\ \citenamefont {Queiroz}}]{Arcadi:2017kky}%
  \BibitemOpen
  \bibfield  {author} {\bibinfo {author} {\bibfnamefont {Giorgio}\ \bibnamefont
  {Arcadi}}, \bibinfo {author} {\bibfnamefont {Ma\'\i{}ra}\ \bibnamefont
  {Dutra}}, \bibinfo {author} {\bibfnamefont {Pradipta}\ \bibnamefont {Ghosh}},
  \bibinfo {author} {\bibfnamefont {Manfred}\ \bibnamefont {Lindner}}, \bibinfo
  {author} {\bibfnamefont {Yann}\ \bibnamefont {Mambrini}}, \bibinfo {author}
  {\bibfnamefont {Mathias}\ \bibnamefont {Pierre}}, \bibinfo {author}
  {\bibfnamefont {Stefano}\ \bibnamefont {Profumo}}, \ and\ \bibinfo {author}
  {\bibfnamefont {Farinaldo~S.}\ \bibnamefont {Queiroz}},\ }\bibfield  {title}
  {\enquote {\bibinfo {title} {{The waning of the WIMP? A review of models,
  searches, and constraints}},}\ }\href {\doibase
  10.1140/epjc/s10052-018-5662-y} {\bibfield  {journal} {\bibinfo  {journal}
  {Eur. Phys. J. C}\ }\textbf {\bibinfo {volume} {78}},\ \bibinfo {pages} {203}
  (\bibinfo {year} {2018})},\ \Eprint {http://arxiv.org/abs/1703.07364}
  {arXiv:1703.07364 [hep-ph]} \BibitemShut {NoStop}%
\bibitem [{\citenamefont {Aalbers}\ \emph {et~al.}(2022)\citenamefont {Aalbers}
  \emph {et~al.}}]{LZ:2022ufs}%
  \BibitemOpen
  \bibfield  {author} {\bibinfo {author} {\bibfnamefont {J.}~\bibnamefont
  {Aalbers}} \emph {et~al.} (\bibinfo {collaboration} {LZ}),\ }\bibfield
  {title} {\enquote {\bibinfo {title} {{First Dark Matter Search Results from
  the LUX-ZEPLIN (LZ) Experiment}},}\ }\href@noop {} {\  (\bibinfo {year}
  {2022})},\ \Eprint {http://arxiv.org/abs/2207.03764} {arXiv:2207.03764
  [hep-ex]} \BibitemShut {NoStop}%
\bibitem [{\citenamefont {Ferreira}(2021)}]{Ferreira:2020fam}%
  \BibitemOpen
  \bibfield  {author} {\bibinfo {author} {\bibfnamefont {Elisa G.~M.}\
  \bibnamefont {Ferreira}},\ }\bibfield  {title} {\enquote {\bibinfo {title}
  {{Ultra-light dark matter}},}\ }\href {\doibase 10.1007/s00159-021-00135-6}
  {\bibfield  {journal} {\bibinfo  {journal} {Astron. Astrophys. Rev.}\
  }\textbf {\bibinfo {volume} {29}},\ \bibinfo {pages} {7} (\bibinfo {year}
  {2021})},\ \Eprint {http://arxiv.org/abs/2005.03254} {arXiv:2005.03254
  [astro-ph.CO]} \BibitemShut {NoStop}%
\bibitem [{\citenamefont {Antypas}\ \emph {et~al.}(2022)\citenamefont {Antypas}
  \emph {et~al.}}]{Antypas:2022asj}%
  \BibitemOpen
  \bibfield  {author} {\bibinfo {author} {\bibfnamefont {D.}~\bibnamefont
  {Antypas}} \emph {et~al.},\ }\bibfield  {title} {\enquote {\bibinfo {title}
  {{New Horizons: Scalar and Vector Ultralight Dark Matter}},}\ }\href@noop {}
  {\  (\bibinfo {year} {2022})},\ \Eprint {http://arxiv.org/abs/2203.14915}
  {arXiv:2203.14915 [hep-ex]} \BibitemShut {NoStop}%
\bibitem [{\citenamefont {Chadha-Day}\ \emph {et~al.}(2022)\citenamefont
  {Chadha-Day}, \citenamefont {Ellis},\ and\ \citenamefont
  {Marsh}}]{Chadha-Day:2021szb}%
  \BibitemOpen
  \bibfield  {author} {\bibinfo {author} {\bibfnamefont {Francesca}\
  \bibnamefont {Chadha-Day}}, \bibinfo {author} {\bibfnamefont {John}\
  \bibnamefont {Ellis}}, \ and\ \bibinfo {author} {\bibfnamefont {David J.~E.}\
  \bibnamefont {Marsh}},\ }\bibfield  {title} {\enquote {\bibinfo {title}
  {{Axion dark matter: What is it and why now?}}}\ }\href {\doibase
  10.1126/sciadv.abj3618} {\bibfield  {journal} {\bibinfo  {journal} {Sci.
  Adv.}\ }\textbf {\bibinfo {volume} {8}},\ \bibinfo {pages} {abj3618}
  (\bibinfo {year} {2022})},\ \Eprint {http://arxiv.org/abs/2105.01406}
  {arXiv:2105.01406 [hep-ph]} \BibitemShut {NoStop}%
\bibitem [{\citenamefont {Peccei}\ and\ \citenamefont
  {Quinn}(1977)}]{Peccei:1977hh}%
  \BibitemOpen
  \bibfield  {author} {\bibinfo {author} {\bibfnamefont {R.~D.}\ \bibnamefont
  {Peccei}}\ and\ \bibinfo {author} {\bibfnamefont {Helen~R.}\ \bibnamefont
  {Quinn}},\ }\bibfield  {title} {\enquote {\bibinfo {title} {{CP Conservation
  in the Presence of Instantons}},}\ }\href {\doibase
  10.1103/PhysRevLett.38.1440} {\bibfield  {journal} {\bibinfo  {journal}
  {Phys. Rev. Lett.}\ }\textbf {\bibinfo {volume} {38}},\ \bibinfo {pages}
  {1440--1443} (\bibinfo {year} {1977})}\BibitemShut {NoStop}%
\bibitem [{\citenamefont {Weinberg}(1978)}]{Weinberg:1977ma}%
  \BibitemOpen
  \bibfield  {author} {\bibinfo {author} {\bibfnamefont {Steven}\ \bibnamefont
  {Weinberg}},\ }\bibfield  {title} {\enquote {\bibinfo {title} {{A New Light
  Boson?}}}\ }\href {\doibase 10.1103/PhysRevLett.40.223} {\bibfield  {journal}
  {\bibinfo  {journal} {Phys. Rev. Lett.}\ }\textbf {\bibinfo {volume} {40}},\
  \bibinfo {pages} {223--226} (\bibinfo {year} {1978})}\BibitemShut {NoStop}%
\bibitem [{\citenamefont {Carney}\ \emph {et~al.}(2022)\citenamefont {Carney}
  \emph {et~al.}}]{Carney:2022gse}%
  \BibitemOpen
  \bibfield  {author} {\bibinfo {author} {\bibfnamefont {Daniel}\ \bibnamefont
  {Carney}} \emph {et~al.},\ }\bibfield  {title} {\enquote {\bibinfo {title}
  {{Snowmass2021 Cosmic Frontier White Paper: Ultraheavy particle dark
  matter}},}\ }\href@noop {} {\  (\bibinfo {year} {2022})},\ \Eprint
  {http://arxiv.org/abs/2203.06508} {arXiv:2203.06508 [hep-ph]} \BibitemShut
  {NoStop}%
\bibitem [{\citenamefont {Bellazzini}\ \emph {et~al.}(2014)\citenamefont
  {Bellazzini}, \citenamefont {Cs\'aki},\ and\ \citenamefont
  {Serra}}]{Bellazzini:2014yua}%
  \BibitemOpen
  \bibfield  {author} {\bibinfo {author} {\bibfnamefont {Brando}\ \bibnamefont
  {Bellazzini}}, \bibinfo {author} {\bibfnamefont {Csaba}\ \bibnamefont
  {Cs\'aki}}, \ and\ \bibinfo {author} {\bibfnamefont {Javi}\ \bibnamefont
  {Serra}},\ }\bibfield  {title} {\enquote {\bibinfo {title} {{Composite
  Higgses}},}\ }\href {\doibase 10.1140/epjc/s10052-014-2766-x} {\bibfield
  {journal} {\bibinfo  {journal} {Eur. Phys. J. C}\ }\textbf {\bibinfo {volume}
  {74}},\ \bibinfo {pages} {2766} (\bibinfo {year} {2014})},\ \Eprint
  {http://arxiv.org/abs/1401.2457} {arXiv:1401.2457 [hep-ph]} \BibitemShut
  {NoStop}%
\bibitem [{\citenamefont {Panico}\ and\ \citenamefont
  {Wulzer}(2016)}]{Panico:2015jxa}%
  \BibitemOpen
  \bibfield  {author} {\bibinfo {author} {\bibfnamefont {Giuliano}\
  \bibnamefont {Panico}}\ and\ \bibinfo {author} {\bibfnamefont {Andrea}\
  \bibnamefont {Wulzer}},\ }\href {\doibase 10.1007/978-3-319-22617-0} {\emph
  {\bibinfo {title} {{The Composite Nambu-Goldstone Higgs}}}},\ Vol.\ \bibinfo
  {volume} {913}\ (\bibinfo  {publisher} {Springer},\ \bibinfo {year} {2016})\
  \Eprint {http://arxiv.org/abs/1506.01961} {arXiv:1506.01961 [hep-ph]}
  \BibitemShut {NoStop}%
\bibitem [{\citenamefont {Hall}\ \emph {et~al.}(2010)\citenamefont {Hall},
  \citenamefont {Jedamzik}, \citenamefont {March-Russell},\ and\ \citenamefont
  {West}}]{Hall:2009bx}%
  \BibitemOpen
  \bibfield  {author} {\bibinfo {author} {\bibfnamefont {Lawrence~J.}\
  \bibnamefont {Hall}}, \bibinfo {author} {\bibfnamefont {Karsten}\
  \bibnamefont {Jedamzik}}, \bibinfo {author} {\bibfnamefont {John}\
  \bibnamefont {March-Russell}}, \ and\ \bibinfo {author} {\bibfnamefont
  {Stephen~M.}\ \bibnamefont {West}},\ }\bibfield  {title} {\enquote {\bibinfo
  {title} {{Freeze-In Production of FIMP Dark Matter}},}\ }\href {\doibase
  10.1007/JHEP03(2010)080} {\bibfield  {journal} {\bibinfo  {journal} {JHEP}\
  }\textbf {\bibinfo {volume} {03}},\ \bibinfo {pages} {080} (\bibinfo {year}
  {2010})},\ \Eprint {http://arxiv.org/abs/0911.1120} {arXiv:0911.1120
  [hep-ph]} \BibitemShut {NoStop}%
\bibitem [{\citenamefont {Elahi}\ \emph {et~al.}(2015)\citenamefont {Elahi},
  \citenamefont {Kolda},\ and\ \citenamefont {Unwin}}]{Elahi:2014fsa}%
  \BibitemOpen
  \bibfield  {author} {\bibinfo {author} {\bibfnamefont {Fatemeh}\ \bibnamefont
  {Elahi}}, \bibinfo {author} {\bibfnamefont {Christopher}\ \bibnamefont
  {Kolda}}, \ and\ \bibinfo {author} {\bibfnamefont {James}\ \bibnamefont
  {Unwin}},\ }\bibfield  {title} {\enquote {\bibinfo {title} {{UltraViolet
  Freeze-in}},}\ }\href {\doibase 10.1007/JHEP03(2015)048} {\bibfield
  {journal} {\bibinfo  {journal} {JHEP}\ }\textbf {\bibinfo {volume} {03}},\
  \bibinfo {pages} {048} (\bibinfo {year} {2015})},\ \Eprint
  {http://arxiv.org/abs/1410.6157} {arXiv:1410.6157 [hep-ph]} \BibitemShut
  {NoStop}%
\bibitem [{\citenamefont {Arkani-Hamed}\ \emph {et~al.}(2001)\citenamefont
  {Arkani-Hamed}, \citenamefont {Porrati},\ and\ \citenamefont
  {Randall}}]{Arkani-Hamed:2000ijo}%
  \BibitemOpen
  \bibfield  {author} {\bibinfo {author} {\bibfnamefont {Nima}\ \bibnamefont
  {Arkani-Hamed}}, \bibinfo {author} {\bibfnamefont {Massimo}\ \bibnamefont
  {Porrati}}, \ and\ \bibinfo {author} {\bibfnamefont {Lisa}\ \bibnamefont
  {Randall}},\ }\bibfield  {title} {\enquote {\bibinfo {title} {{Holography and
  phenomenology}},}\ }\href {\doibase 10.1088/1126-6708/2001/08/017} {\bibfield
   {journal} {\bibinfo  {journal} {JHEP}\ }\textbf {\bibinfo {volume} {08}},\
  \bibinfo {pages} {017} (\bibinfo {year} {2001})},\ \Eprint
  {http://arxiv.org/abs/hep-th/0012148} {arXiv:hep-th/0012148} \BibitemShut
  {NoStop}%
\bibitem [{\citenamefont {Rattazzi}\ and\ \citenamefont
  {Zaffaroni}(2001)}]{Rattazzi:2000hs}%
  \BibitemOpen
  \bibfield  {author} {\bibinfo {author} {\bibfnamefont {R.}~\bibnamefont
  {Rattazzi}}\ and\ \bibinfo {author} {\bibfnamefont {A.}~\bibnamefont
  {Zaffaroni}},\ }\bibfield  {title} {\enquote {\bibinfo {title} {{Comments on
  the holographic picture of the Randall-Sundrum model}},}\ }\href {\doibase
  10.1088/1126-6708/2001/04/021} {\bibfield  {journal} {\bibinfo  {journal}
  {JHEP}\ }\textbf {\bibinfo {volume} {04}},\ \bibinfo {pages} {021} (\bibinfo
  {year} {2001})},\ \Eprint {http://arxiv.org/abs/hep-th/0012248}
  {arXiv:hep-th/0012248} \BibitemShut {NoStop}%
\bibitem [{\citenamefont {Randall}\ and\ \citenamefont
  {Sundrum}(1999)}]{Randall:1999ee}%
  \BibitemOpen
  \bibfield  {author} {\bibinfo {author} {\bibfnamefont {Lisa}\ \bibnamefont
  {Randall}}\ and\ \bibinfo {author} {\bibfnamefont {Raman}\ \bibnamefont
  {Sundrum}},\ }\bibfield  {title} {\enquote {\bibinfo {title} {{A Large mass
  hierarchy from a small extra dimension}},}\ }\href {\doibase
  10.1103/PhysRevLett.83.3370} {\bibfield  {journal} {\bibinfo  {journal}
  {Phys. Rev. Lett.}\ }\textbf {\bibinfo {volume} {83}},\ \bibinfo {pages}
  {3370--3373} (\bibinfo {year} {1999})},\ \Eprint
  {http://arxiv.org/abs/hep-ph/9905221} {arXiv:hep-ph/9905221} \BibitemShut
  {NoStop}%
\bibitem [{\citenamefont {Contino}\ \emph {et~al.}(2003)\citenamefont
  {Contino}, \citenamefont {Nomura},\ and\ \citenamefont
  {Pomarol}}]{Contino:2003ve}%
  \BibitemOpen
  \bibfield  {author} {\bibinfo {author} {\bibfnamefont {Roberto}\ \bibnamefont
  {Contino}}, \bibinfo {author} {\bibfnamefont {Yasunori}\ \bibnamefont
  {Nomura}}, \ and\ \bibinfo {author} {\bibfnamefont {Alex}\ \bibnamefont
  {Pomarol}},\ }\bibfield  {title} {\enquote {\bibinfo {title} {{Higgs as a
  holographic pseudoGoldstone boson}},}\ }\href {\doibase
  10.1016/j.nuclphysb.2003.08.027} {\bibfield  {journal} {\bibinfo  {journal}
  {Nucl. Phys. B}\ }\textbf {\bibinfo {volume} {671}},\ \bibinfo {pages}
  {148--174} (\bibinfo {year} {2003})},\ \Eprint
  {http://arxiv.org/abs/hep-ph/0306259} {arXiv:hep-ph/0306259} \BibitemShut
  {NoStop}%
\bibitem [{\citenamefont {Erdmenger}\ \emph {et~al.}(2021)\citenamefont
  {Erdmenger}, \citenamefont {Evans}, \citenamefont {Porod},\ and\
  \citenamefont {Rigatos}}]{Erdmenger:2020lvq}%
  \BibitemOpen
  \bibfield  {author} {\bibinfo {author} {\bibfnamefont {Johanna}\ \bibnamefont
  {Erdmenger}}, \bibinfo {author} {\bibfnamefont {Nick}\ \bibnamefont {Evans}},
  \bibinfo {author} {\bibfnamefont {Werner}\ \bibnamefont {Porod}}, \ and\
  \bibinfo {author} {\bibfnamefont {Konstantinos~S.}\ \bibnamefont {Rigatos}},\
  }\bibfield  {title} {\enquote {\bibinfo {title} {{Gauge/gravity dynamics for
  composite Higgs models and the top mass}},}\ }\href {\doibase
  10.1103/PhysRevLett.126.071602} {\bibfield  {journal} {\bibinfo  {journal}
  {Phys. Rev. Lett.}\ }\textbf {\bibinfo {volume} {126}},\ \bibinfo {pages}
  {071602} (\bibinfo {year} {2021})},\ \Eprint
  {http://arxiv.org/abs/2009.10737} {arXiv:2009.10737 [hep-ph]} \BibitemShut
  {NoStop}%
\bibitem [{\citenamefont {Bernal}\ \emph {et~al.}(2017)\citenamefont {Bernal},
  \citenamefont {Heikinheimo}, \citenamefont {Tenkanen}, \citenamefont
  {Tuominen},\ and\ \citenamefont {Vaskonen}}]{Bernal:2017kxu}%
  \BibitemOpen
  \bibfield  {author} {\bibinfo {author} {\bibfnamefont {Nicol\'as}\
  \bibnamefont {Bernal}}, \bibinfo {author} {\bibfnamefont {Matti}\
  \bibnamefont {Heikinheimo}}, \bibinfo {author} {\bibfnamefont {Tommi}\
  \bibnamefont {Tenkanen}}, \bibinfo {author} {\bibfnamefont {Kimmo}\
  \bibnamefont {Tuominen}}, \ and\ \bibinfo {author} {\bibfnamefont {Ville}\
  \bibnamefont {Vaskonen}},\ }\bibfield  {title} {\enquote {\bibinfo {title}
  {{The Dawn of FIMP Dark Matter: A Review of Models and Constraints}},}\
  }\href {\doibase 10.1142/S0217751X1730023X} {\bibfield  {journal} {\bibinfo
  {journal} {Int. J. Mod. Phys. A}\ }\textbf {\bibinfo {volume} {32}},\
  \bibinfo {pages} {1730023} (\bibinfo {year} {2017})},\ \Eprint
  {http://arxiv.org/abs/1706.07442} {arXiv:1706.07442 [hep-ph]} \BibitemShut
  {NoStop}%
\bibitem [{\citenamefont {Hong}\ \emph {et~al.}(2020)\citenamefont {Hong},
  \citenamefont {Kurup},\ and\ \citenamefont {Perelstein}}]{Hong:2019nwd}%
  \BibitemOpen
  \bibfield  {author} {\bibinfo {author} {\bibfnamefont {Sungwoo}\ \bibnamefont
  {Hong}}, \bibinfo {author} {\bibfnamefont {Gowri}\ \bibnamefont {Kurup}}, \
  and\ \bibinfo {author} {\bibfnamefont {Maxim}\ \bibnamefont {Perelstein}},\
  }\bibfield  {title} {\enquote {\bibinfo {title} {{Conformal Freeze-In of Dark
  Matter}},}\ }\href {\doibase 10.1103/PhysRevD.101.095037} {\bibfield
  {journal} {\bibinfo  {journal} {Phys. Rev. D}\ }\textbf {\bibinfo {volume}
  {101}},\ \bibinfo {pages} {095037} (\bibinfo {year} {2020})},\ \Eprint
  {http://arxiv.org/abs/1910.10160} {arXiv:1910.10160 [hep-ph]} \BibitemShut
  {NoStop}%
\bibitem [{\citenamefont {Hong}\ \emph {et~al.}(2023)\citenamefont {Hong},
  \citenamefont {Kurup},\ and\ \citenamefont {Perelstein}}]{Hong:2022gzo}%
  \BibitemOpen
  \bibfield  {author} {\bibinfo {author} {\bibfnamefont {Sungwoo}\ \bibnamefont
  {Hong}}, \bibinfo {author} {\bibfnamefont {Gowri}\ \bibnamefont {Kurup}}, \
  and\ \bibinfo {author} {\bibfnamefont {Maxim}\ \bibnamefont {Perelstein}},\
  }\bibfield  {title} {\enquote {\bibinfo {title} {{Dark matter from a
  conformal Dark Sector}},}\ }\href {\doibase 10.1007/JHEP02(2023)221}
  {\bibfield  {journal} {\bibinfo  {journal} {JHEP}\ }\textbf {\bibinfo
  {volume} {02}},\ \bibinfo {pages} {221} (\bibinfo {year} {2023})},\ \Eprint
  {http://arxiv.org/abs/2207.10093} {arXiv:2207.10093 [hep-ph]} \BibitemShut
  {NoStop}%
\bibitem [{\citenamefont {Chiu}\ \emph {et~al.}(2023)\citenamefont {Chiu},
  \citenamefont {Hong},\ and\ \citenamefont {Wang}}]{Chiu:2022bni}%
  \BibitemOpen
  \bibfield  {author} {\bibinfo {author} {\bibfnamefont {Wen~Han}\ \bibnamefont
  {Chiu}}, \bibinfo {author} {\bibfnamefont {Sungwoo}\ \bibnamefont {Hong}}, \
  and\ \bibinfo {author} {\bibfnamefont {Lian-Tao}\ \bibnamefont {Wang}},\
  }\bibfield  {title} {\enquote {\bibinfo {title} {{Conformal freeze-in,
  composite dark photon, and asymmetric reheating}},}\ }\href {\doibase
  10.1007/JHEP03(2023)172} {\bibfield  {journal} {\bibinfo  {journal} {JHEP}\
  }\textbf {\bibinfo {volume} {03}},\ \bibinfo {pages} {172} (\bibinfo {year}
  {2023})},\ \Eprint {http://arxiv.org/abs/2209.10563} {arXiv:2209.10563
  [hep-ph]} \BibitemShut {NoStop}%
\bibitem [{\citenamefont {Blum}\ \emph {et~al.}(2015)\citenamefont {Blum},
  \citenamefont {Cliche}, \citenamefont {Csaki},\ and\ \citenamefont
  {Lee}}]{Blum:2014jca}%
  \BibitemOpen
  \bibfield  {author} {\bibinfo {author} {\bibfnamefont {Kfir}\ \bibnamefont
  {Blum}}, \bibinfo {author} {\bibfnamefont {Mathieu}\ \bibnamefont {Cliche}},
  \bibinfo {author} {\bibfnamefont {Csaba}\ \bibnamefont {Csaki}}, \ and\
  \bibinfo {author} {\bibfnamefont {Seung~J.}\ \bibnamefont {Lee}},\ }\bibfield
   {title} {\enquote {\bibinfo {title} {{WIMP Dark Matter through the Dilaton
  Portal}},}\ }\href {\doibase 10.1007/JHEP03(2015)099} {\bibfield  {journal}
  {\bibinfo  {journal} {JHEP}\ }\textbf {\bibinfo {volume} {03}},\ \bibinfo
  {pages} {099} (\bibinfo {year} {2015})},\ \Eprint
  {http://arxiv.org/abs/1410.1873} {arXiv:1410.1873 [hep-ph]} \BibitemShut
  {NoStop}%
\bibitem [{\citenamefont {von Harling}\ and\ \citenamefont
  {McDonald}(2012)}]{vonHarling:2012sz}%
  \BibitemOpen
  \bibfield  {author} {\bibinfo {author} {\bibfnamefont {Benedict}\
  \bibnamefont {von Harling}}\ and\ \bibinfo {author} {\bibfnamefont
  {Kristian~L.}\ \bibnamefont {McDonald}},\ }\bibfield  {title} {\enquote
  {\bibinfo {title} {{Secluded Dark Matter Coupled to a Hidden CFT}},}\ }\href
  {\doibase 10.1007/JHEP08(2012)048} {\bibfield  {journal} {\bibinfo  {journal}
  {JHEP}\ }\textbf {\bibinfo {volume} {08}},\ \bibinfo {pages} {048} (\bibinfo
  {year} {2012})},\ \Eprint {http://arxiv.org/abs/1203.6646} {arXiv:1203.6646
  [hep-ph]} \BibitemShut {NoStop}%
\bibitem [{\citenamefont {McDonald}(2012)}]{McDonald:2012nc}%
  \BibitemOpen
  \bibfield  {author} {\bibinfo {author} {\bibfnamefont {Kristian~L.}\
  \bibnamefont {McDonald}},\ }\bibfield  {title} {\enquote {\bibinfo {title}
  {{Sommerfeld Enhancement from Multiple Mediators}},}\ }\href {\doibase
  10.1007/JHEP07(2012)145} {\bibfield  {journal} {\bibinfo  {journal} {JHEP}\
  }\textbf {\bibinfo {volume} {07}},\ \bibinfo {pages} {145} (\bibinfo {year}
  {2012})},\ \Eprint {http://arxiv.org/abs/1203.6341} {arXiv:1203.6341
  [hep-ph]} \BibitemShut {NoStop}%
\bibitem [{\citenamefont {McDonald}\ and\ \citenamefont
  {Morrissey}(2011)}]{McDonald:2010fe}%
  \BibitemOpen
  \bibfield  {author} {\bibinfo {author} {\bibfnamefont {Kristian~L.}\
  \bibnamefont {McDonald}}\ and\ \bibinfo {author} {\bibfnamefont {David~E.}\
  \bibnamefont {Morrissey}},\ }\bibfield  {title} {\enquote {\bibinfo {title}
  {{Low-Energy Signals from Kinetic Mixing with a Warped Abelian Hidden
  Sector}},}\ }\href {\doibase 10.1007/JHEP02(2011)087} {\bibfield  {journal}
  {\bibinfo  {journal} {JHEP}\ }\textbf {\bibinfo {volume} {02}},\ \bibinfo
  {pages} {087} (\bibinfo {year} {2011})},\ \Eprint
  {http://arxiv.org/abs/1010.5999} {arXiv:1010.5999 [hep-ph]} \BibitemShut
  {NoStop}%
\bibitem [{\citenamefont {McDonald}\ and\ \citenamefont
  {Morrissey}(2010)}]{McDonald:2010iq}%
  \BibitemOpen
  \bibfield  {author} {\bibinfo {author} {\bibfnamefont {Kristian~L.}\
  \bibnamefont {McDonald}}\ and\ \bibinfo {author} {\bibfnamefont {David~E.}\
  \bibnamefont {Morrissey}},\ }\bibfield  {title} {\enquote {\bibinfo {title}
  {{Low-Energy Probes of a Warped Extra Dimension}},}\ }\href {\doibase
  10.1007/JHEP05(2010)056} {\bibfield  {journal} {\bibinfo  {journal} {JHEP}\
  }\textbf {\bibinfo {volume} {05}},\ \bibinfo {pages} {056} (\bibinfo {year}
  {2010})},\ \Eprint {http://arxiv.org/abs/1002.3361} {arXiv:1002.3361
  [hep-ph]} \BibitemShut {NoStop}%
\bibitem [{\citenamefont {Brax}\ \emph {et~al.}(2019)\citenamefont {Brax},
  \citenamefont {Fichet},\ and\ \citenamefont {Tanedo}}]{Brax:2019koq}%
  \BibitemOpen
  \bibfield  {author} {\bibinfo {author} {\bibfnamefont {Philippe}\
  \bibnamefont {Brax}}, \bibinfo {author} {\bibfnamefont {Sylvain}\
  \bibnamefont {Fichet}}, \ and\ \bibinfo {author} {\bibfnamefont {Philip}\
  \bibnamefont {Tanedo}},\ }\bibfield  {title} {\enquote {\bibinfo {title}
  {{The Warped Dark Sector}},}\ }\href {\doibase
  10.1016/j.physletb.2019.135012} {\bibfield  {journal} {\bibinfo  {journal}
  {Phys. Lett. B}\ }\textbf {\bibinfo {volume} {798}},\ \bibinfo {pages}
  {135012} (\bibinfo {year} {2019})},\ \Eprint
  {http://arxiv.org/abs/1906.02199} {arXiv:1906.02199 [hep-ph]} \BibitemShut
  {NoStop}%
\bibitem [{\citenamefont {Cs\'aki}\ \emph
  {et~al.}(2022{\natexlab{a}})\citenamefont {Cs\'aki}, \citenamefont {Hong},
  \citenamefont {Kurup}, \citenamefont {Lee}, \citenamefont {Perelstein},\ and\
  \citenamefont {Xue}}]{Csaki:2021gfm}%
  \BibitemOpen
  \bibfield  {author} {\bibinfo {author} {\bibfnamefont {Csaba}\ \bibnamefont
  {Cs\'aki}}, \bibinfo {author} {\bibfnamefont {Sungwoo}\ \bibnamefont {Hong}},
  \bibinfo {author} {\bibfnamefont {Gowri}\ \bibnamefont {Kurup}}, \bibinfo
  {author} {\bibfnamefont {Seung~J.}\ \bibnamefont {Lee}}, \bibinfo {author}
  {\bibfnamefont {Maxim}\ \bibnamefont {Perelstein}}, \ and\ \bibinfo {author}
  {\bibfnamefont {Wei}\ \bibnamefont {Xue}},\ }\bibfield  {title} {\enquote
  {\bibinfo {title} {{Continuum dark matter}},}\ }\href {\doibase
  10.1103/PhysRevD.105.035025} {\bibfield  {journal} {\bibinfo  {journal}
  {Phys. Rev. D}\ }\textbf {\bibinfo {volume} {105}},\ \bibinfo {pages}
  {035025} (\bibinfo {year} {2022}{\natexlab{a}})},\ \Eprint
  {http://arxiv.org/abs/2105.07035} {arXiv:2105.07035 [hep-ph]} \BibitemShut
  {NoStop}%
\bibitem [{\citenamefont {Cs\'aki}\ \emph
  {et~al.}(2022{\natexlab{b}})\citenamefont {Cs\'aki}, \citenamefont {Hong},
  \citenamefont {Kurup}, \citenamefont {Lee}, \citenamefont {Perelstein},\ and\
  \citenamefont {Xue}}]{Csaki:2021xpy}%
  \BibitemOpen
  \bibfield  {author} {\bibinfo {author} {\bibfnamefont {Csaba}\ \bibnamefont
  {Cs\'aki}}, \bibinfo {author} {\bibfnamefont {Sungwoo}\ \bibnamefont {Hong}},
  \bibinfo {author} {\bibfnamefont {Gowri}\ \bibnamefont {Kurup}}, \bibinfo
  {author} {\bibfnamefont {Seung~J.}\ \bibnamefont {Lee}}, \bibinfo {author}
  {\bibfnamefont {Maxim}\ \bibnamefont {Perelstein}}, \ and\ \bibinfo {author}
  {\bibfnamefont {Wei}\ \bibnamefont {Xue}},\ }\bibfield  {title} {\enquote
  {\bibinfo {title} {{Z-Portal Continuum Dark Matter}},}\ }\href {\doibase
  10.1103/PhysRevLett.128.081807} {\bibfield  {journal} {\bibinfo  {journal}
  {Phys. Rev. Lett.}\ }\textbf {\bibinfo {volume} {128}},\ \bibinfo {pages}
  {081807} (\bibinfo {year} {2022}{\natexlab{b}})},\ \Eprint
  {http://arxiv.org/abs/2105.14023} {arXiv:2105.14023 [hep-ph]} \BibitemShut
  {NoStop}%
\bibitem [{\citenamefont {Csaki}\ \emph {et~al.}(2023)\citenamefont {Csaki},
  \citenamefont {Ismail},\ and\ \citenamefont {Lee}}]{Csaki:2022lnq}%
  \BibitemOpen
  \bibfield  {author} {\bibinfo {author} {\bibfnamefont {Csaba}\ \bibnamefont
  {Csaki}}, \bibinfo {author} {\bibfnamefont {Ameen}\ \bibnamefont {Ismail}}, \
  and\ \bibinfo {author} {\bibfnamefont {Seung~J.}\ \bibnamefont {Lee}},\
  }\bibfield  {title} {\enquote {\bibinfo {title} {{The continuum dark matter
  zoo}},}\ }\href {\doibase 10.1007/JHEP02(2023)053} {\bibfield  {journal}
  {\bibinfo  {journal} {JHEP}\ }\textbf {\bibinfo {volume} {02}},\ \bibinfo
  {pages} {053} (\bibinfo {year} {2023})},\ \Eprint
  {http://arxiv.org/abs/2210.16326} {arXiv:2210.16326 [hep-ph]} \BibitemShut
  {NoStop}%
\bibitem [{\citenamefont {Fichet}\ \emph
  {et~al.}(2023{\natexlab{a}})\citenamefont {Fichet}, \citenamefont {Megias},\
  and\ \citenamefont {Quiros}}]{Fichet:2022ixi}%
  \BibitemOpen
  \bibfield  {author} {\bibinfo {author} {\bibfnamefont {Sylvain}\ \bibnamefont
  {Fichet}}, \bibinfo {author} {\bibfnamefont {Eugenio}\ \bibnamefont
  {Megias}}, \ and\ \bibinfo {author} {\bibfnamefont {Mariano}\ \bibnamefont
  {Quiros}},\ }\bibfield  {title} {\enquote {\bibinfo {title} {{Continuum
  effective field theories, gravity, and holography}},}\ }\href {\doibase
  10.1103/PhysRevD.107.096016} {\bibfield  {journal} {\bibinfo  {journal}
  {Phys. Rev. D}\ }\textbf {\bibinfo {volume} {107}},\ \bibinfo {pages}
  {096016} (\bibinfo {year} {2023}{\natexlab{a}})},\ \Eprint
  {http://arxiv.org/abs/2208.12273} {arXiv:2208.12273 [hep-ph]} \BibitemShut
  {NoStop}%
\bibitem [{\citenamefont {Fichet}\ \emph
  {et~al.}(2023{\natexlab{b}})\citenamefont {Fichet}, \citenamefont {Megias},\
  and\ \citenamefont {Quiros}}]{Fichet:2022xol}%
  \BibitemOpen
  \bibfield  {author} {\bibinfo {author} {\bibfnamefont {Sylvain}\ \bibnamefont
  {Fichet}}, \bibinfo {author} {\bibfnamefont {Eugenio}\ \bibnamefont
  {Megias}}, \ and\ \bibinfo {author} {\bibfnamefont {Mariano}\ \bibnamefont
  {Quiros}},\ }\bibfield  {title} {\enquote {\bibinfo {title} {{Cosmological
  dark matter from a bulk black hole}},}\ }\href {\doibase
  10.1103/PhysRevD.107.115014} {\bibfield  {journal} {\bibinfo  {journal}
  {Phys. Rev. D}\ }\textbf {\bibinfo {volume} {107}},\ \bibinfo {pages}
  {115014} (\bibinfo {year} {2023}{\natexlab{b}})},\ \Eprint
  {http://arxiv.org/abs/2212.13268} {arXiv:2212.13268 [hep-ph]} \BibitemShut
  {NoStop}%
\bibitem [{\citenamefont {Bernal}\ \emph {et~al.}(2020)\citenamefont {Bernal},
  \citenamefont {Donini}, \citenamefont {Folgado},\ and\ \citenamefont
  {Rius}}]{Bernal:2020fvw}%
  \BibitemOpen
  \bibfield  {author} {\bibinfo {author} {\bibfnamefont {Nicol\'as}\
  \bibnamefont {Bernal}}, \bibinfo {author} {\bibfnamefont {Andrea}\
  \bibnamefont {Donini}}, \bibinfo {author} {\bibfnamefont {Miguel~G.}\
  \bibnamefont {Folgado}}, \ and\ \bibinfo {author} {\bibfnamefont {Nuria}\
  \bibnamefont {Rius}},\ }\bibfield  {title} {\enquote {\bibinfo {title}
  {{Kaluza-Klein FIMP Dark Matter in Warped Extra-Dimensions}},}\ }\href
  {\doibase 10.1007/JHEP09(2020)142} {\bibfield  {journal} {\bibinfo  {journal}
  {JHEP}\ }\textbf {\bibinfo {volume} {09}},\ \bibinfo {pages} {142} (\bibinfo
  {year} {2020})},\ \Eprint {http://arxiv.org/abs/2004.14403} {arXiv:2004.14403
  [hep-ph]} \BibitemShut {NoStop}%
\bibitem [{\citenamefont {Bernal}\ \emph {et~al.}(2021)\citenamefont {Bernal},
  \citenamefont {Donini}, \citenamefont {Folgado},\ and\ \citenamefont
  {Rius}}]{Bernal:2020yqg}%
  \BibitemOpen
  \bibfield  {author} {\bibinfo {author} {\bibfnamefont {Nicol\'as}\
  \bibnamefont {Bernal}}, \bibinfo {author} {\bibfnamefont {Andrea}\
  \bibnamefont {Donini}}, \bibinfo {author} {\bibfnamefont {Miguel~G.}\
  \bibnamefont {Folgado}}, \ and\ \bibinfo {author} {\bibfnamefont {Nuria}\
  \bibnamefont {Rius}},\ }\bibfield  {title} {\enquote {\bibinfo {title} {{FIMP
  Dark Matter in Clockwork/Linear Dilaton Extra-Dimensions}},}\ }\href
  {\doibase 10.1007/JHEP04(2021)061} {\bibfield  {journal} {\bibinfo  {journal}
  {JHEP}\ }\textbf {\bibinfo {volume} {04}},\ \bibinfo {pages} {061} (\bibinfo
  {year} {2021})},\ \Eprint {http://arxiv.org/abs/2012.10453} {arXiv:2012.10453
  [hep-ph]} \BibitemShut {NoStop}%
\bibitem [{\citenamefont {de~Giorgi}\ and\ \citenamefont
  {Vogl}(2021)}]{deGiorgi:2021xvm}%
  \BibitemOpen
  \bibfield  {author} {\bibinfo {author} {\bibfnamefont {Arturo}\ \bibnamefont
  {de~Giorgi}}\ and\ \bibinfo {author} {\bibfnamefont {Stefan}\ \bibnamefont
  {Vogl}},\ }\bibfield  {title} {\enquote {\bibinfo {title} {{Dark matter
  interacting via a massive spin-2 mediator in warped extra-dimensions}},}\
  }\href {\doibase 10.1007/JHEP11(2021)036} {\bibfield  {journal} {\bibinfo
  {journal} {JHEP}\ }\textbf {\bibinfo {volume} {11}},\ \bibinfo {pages} {036}
  (\bibinfo {year} {2021})},\ \Eprint {http://arxiv.org/abs/2105.06794}
  {arXiv:2105.06794 [hep-ph]} \BibitemShut {NoStop}%
\bibitem [{\citenamefont {de~Giorgi}\ and\ \citenamefont
  {Vogl}(2023)}]{deGiorgi:2022yha}%
  \BibitemOpen
  \bibfield  {author} {\bibinfo {author} {\bibfnamefont {Arturo}\ \bibnamefont
  {de~Giorgi}}\ and\ \bibinfo {author} {\bibfnamefont {Stefan}\ \bibnamefont
  {Vogl}},\ }\bibfield  {title} {\enquote {\bibinfo {title} {{Warm dark matter
  from a gravitational freeze-in in extra dimensions}},}\ }\href {\doibase
  10.1007/JHEP04(2023)032} {\bibfield  {journal} {\bibinfo  {journal} {JHEP}\
  }\textbf {\bibinfo {volume} {04}},\ \bibinfo {pages} {032} (\bibinfo {year}
  {2023})},\ \Eprint {http://arxiv.org/abs/2208.03153} {arXiv:2208.03153
  [hep-ph]} \BibitemShut {NoStop}%
\bibitem [{\citenamefont {Frigerio}\ \emph {et~al.}(2012)\citenamefont
  {Frigerio}, \citenamefont {Pomarol}, \citenamefont {Riva},\ and\
  \citenamefont {Urbano}}]{Frigerio:2012uc}%
  \BibitemOpen
  \bibfield  {author} {\bibinfo {author} {\bibfnamefont {Michele}\ \bibnamefont
  {Frigerio}}, \bibinfo {author} {\bibfnamefont {Alex}\ \bibnamefont
  {Pomarol}}, \bibinfo {author} {\bibfnamefont {Francesco}\ \bibnamefont
  {Riva}}, \ and\ \bibinfo {author} {\bibfnamefont {Alfredo}\ \bibnamefont
  {Urbano}},\ }\bibfield  {title} {\enquote {\bibinfo {title} {{Composite
  Scalar Dark Matter}},}\ }\href {\doibase 10.1007/JHEP07(2012)015} {\bibfield
  {journal} {\bibinfo  {journal} {JHEP}\ }\textbf {\bibinfo {volume} {07}},\
  \bibinfo {pages} {015} (\bibinfo {year} {2012})},\ \Eprint
  {http://arxiv.org/abs/1204.2808} {arXiv:1204.2808 [hep-ph]} \BibitemShut
  {NoStop}%
\bibitem [{\citenamefont {Chala}(2013)}]{Chala:2012af}%
  \BibitemOpen
  \bibfield  {author} {\bibinfo {author} {\bibfnamefont {Mikael}\ \bibnamefont
  {Chala}},\ }\bibfield  {title} {\enquote {\bibinfo {title} {{$h \rightarrow
  \gamma\gamma$ excess and Dark Matter from Composite Higgs Models}},}\ }\href
  {\doibase 10.1007/JHEP01(2013)122} {\bibfield  {journal} {\bibinfo  {journal}
  {JHEP}\ }\textbf {\bibinfo {volume} {01}},\ \bibinfo {pages} {122} (\bibinfo
  {year} {2013})},\ \Eprint {http://arxiv.org/abs/1210.6208} {arXiv:1210.6208
  [hep-ph]} \BibitemShut {NoStop}%
\bibitem [{\citenamefont {Marzocca}\ and\ \citenamefont
  {Urbano}(2014)}]{Marzocca:2014msa}%
  \BibitemOpen
  \bibfield  {author} {\bibinfo {author} {\bibfnamefont {David}\ \bibnamefont
  {Marzocca}}\ and\ \bibinfo {author} {\bibfnamefont {Alfredo}\ \bibnamefont
  {Urbano}},\ }\bibfield  {title} {\enquote {\bibinfo {title} {{Composite Dark
  Matter and LHC Interplay}},}\ }\href {\doibase 10.1007/JHEP07(2014)107}
  {\bibfield  {journal} {\bibinfo  {journal} {JHEP}\ }\textbf {\bibinfo
  {volume} {07}},\ \bibinfo {pages} {107} (\bibinfo {year} {2014})},\ \Eprint
  {http://arxiv.org/abs/1404.7419} {arXiv:1404.7419 [hep-ph]} \BibitemShut
  {NoStop}%
\bibitem [{\citenamefont {Fonseca}\ \emph {et~al.}(2015)\citenamefont
  {Fonseca}, \citenamefont {Zukanovich~Funchal}, \citenamefont {Lessa},\ and\
  \citenamefont {Lopez-Honorez}}]{Fonseca:2015gva}%
  \BibitemOpen
  \bibfield  {author} {\bibinfo {author} {\bibfnamefont {Nayara}\ \bibnamefont
  {Fonseca}}, \bibinfo {author} {\bibfnamefont {Renata}\ \bibnamefont
  {Zukanovich~Funchal}}, \bibinfo {author} {\bibfnamefont {Andre}\ \bibnamefont
  {Lessa}}, \ and\ \bibinfo {author} {\bibfnamefont {Laura}\ \bibnamefont
  {Lopez-Honorez}},\ }\bibfield  {title} {\enquote {\bibinfo {title} {{Dark
  Matter Constraints on Composite Higgs Models}},}\ }\href {\doibase
  10.1007/JHEP06(2015)154} {\bibfield  {journal} {\bibinfo  {journal} {JHEP}\
  }\textbf {\bibinfo {volume} {06}},\ \bibinfo {pages} {154} (\bibinfo {year}
  {2015})},\ \Eprint {http://arxiv.org/abs/1501.05957} {arXiv:1501.05957
  [hep-ph]} \BibitemShut {NoStop}%
\bibitem [{\citenamefont {Kim}\ \emph {et~al.}(2016)\citenamefont {Kim},
  \citenamefont {Lee},\ and\ \citenamefont {Parolini}}]{Kim:2016jbz}%
  \BibitemOpen
  \bibfield  {author} {\bibinfo {author} {\bibfnamefont {Manki}\ \bibnamefont
  {Kim}}, \bibinfo {author} {\bibfnamefont {Seung~J.}\ \bibnamefont {Lee}}, \
  and\ \bibinfo {author} {\bibfnamefont {Alberto}\ \bibnamefont {Parolini}},\
  }\bibfield  {title} {\enquote {\bibinfo {title} {{WIMP Dark Matter in
  Composite Higgs Models and the Dilaton Portal}},}\ }\href@noop {} {\
  (\bibinfo {year} {2016})},\ \Eprint {http://arxiv.org/abs/1602.05590}
  {arXiv:1602.05590 [hep-ph]} \BibitemShut {NoStop}%
\bibitem [{\citenamefont {Wu}\ \emph {et~al.}(2017)\citenamefont {Wu},
  \citenamefont {Ma}, \citenamefont {Zhang},\ and\ \citenamefont
  {Cacciapaglia}}]{Wu:2017iji}%
  \BibitemOpen
  \bibfield  {author} {\bibinfo {author} {\bibfnamefont {Yongcheng}\
  \bibnamefont {Wu}}, \bibinfo {author} {\bibfnamefont {Teng}\ \bibnamefont
  {Ma}}, \bibinfo {author} {\bibfnamefont {Bin}\ \bibnamefont {Zhang}}, \ and\
  \bibinfo {author} {\bibfnamefont {Giacomo}\ \bibnamefont {Cacciapaglia}},\
  }\bibfield  {title} {\enquote {\bibinfo {title} {{Composite Dark Matter and
  Higgs}},}\ }\href {\doibase 10.1007/JHEP11(2017)058} {\bibfield  {journal}
  {\bibinfo  {journal} {JHEP}\ }\textbf {\bibinfo {volume} {11}},\ \bibinfo
  {pages} {058} (\bibinfo {year} {2017})},\ \Eprint
  {http://arxiv.org/abs/1703.06903} {arXiv:1703.06903 [hep-ph]} \BibitemShut
  {NoStop}%
\bibitem [{\citenamefont {Ballesteros}\ \emph {et~al.}(2017)\citenamefont
  {Ballesteros}, \citenamefont {Carmona},\ and\ \citenamefont
  {Chala}}]{Ballesteros:2017xeg}%
  \BibitemOpen
  \bibfield  {author} {\bibinfo {author} {\bibfnamefont {Guillermo}\
  \bibnamefont {Ballesteros}}, \bibinfo {author} {\bibfnamefont {Adrian}\
  \bibnamefont {Carmona}}, \ and\ \bibinfo {author} {\bibfnamefont {Mikael}\
  \bibnamefont {Chala}},\ }\bibfield  {title} {\enquote {\bibinfo {title}
  {{Exceptional Composite Dark Matter}},}\ }\href {\doibase
  10.1140/epjc/s10052-017-5040-1} {\bibfield  {journal} {\bibinfo  {journal}
  {Eur. Phys. J. C}\ }\textbf {\bibinfo {volume} {77}},\ \bibinfo {pages} {468}
  (\bibinfo {year} {2017})},\ \Eprint {http://arxiv.org/abs/1704.07388}
  {arXiv:1704.07388 [hep-ph]} \BibitemShut {NoStop}%
\bibitem [{\citenamefont {Balkin}\ \emph {et~al.}(2017)\citenamefont {Balkin},
  \citenamefont {Ruhdorfer}, \citenamefont {Salvioni},\ and\ \citenamefont
  {Weiler}}]{Balkin:2017aep}%
  \BibitemOpen
  \bibfield  {author} {\bibinfo {author} {\bibfnamefont {Reuven}\ \bibnamefont
  {Balkin}}, \bibinfo {author} {\bibfnamefont {Maximilian}\ \bibnamefont
  {Ruhdorfer}}, \bibinfo {author} {\bibfnamefont {Ennio}\ \bibnamefont
  {Salvioni}}, \ and\ \bibinfo {author} {\bibfnamefont {Andreas}\ \bibnamefont
  {Weiler}},\ }\bibfield  {title} {\enquote {\bibinfo {title} {{Charged
  Composite Scalar Dark Matter}},}\ }\href {\doibase 10.1007/JHEP11(2017)094}
  {\bibfield  {journal} {\bibinfo  {journal} {JHEP}\ }\textbf {\bibinfo
  {volume} {11}},\ \bibinfo {pages} {094} (\bibinfo {year} {2017})},\ \Eprint
  {http://arxiv.org/abs/1707.07685} {arXiv:1707.07685 [hep-ph]} \BibitemShut
  {NoStop}%
\bibitem [{\citenamefont {Alanne}\ \emph {et~al.}(2018)\citenamefont {Alanne},
  \citenamefont {Buarque~Franzosi}, \citenamefont {Frandsen},\ and\
  \citenamefont {Rosenlyst}}]{Alanne:2018xli}%
  \BibitemOpen
  \bibfield  {author} {\bibinfo {author} {\bibfnamefont {Tommi}\ \bibnamefont
  {Alanne}}, \bibinfo {author} {\bibfnamefont {Diogo}\ \bibnamefont
  {Buarque~Franzosi}}, \bibinfo {author} {\bibfnamefont {Mads~T.}\ \bibnamefont
  {Frandsen}}, \ and\ \bibinfo {author} {\bibfnamefont {Martin}\ \bibnamefont
  {Rosenlyst}},\ }\bibfield  {title} {\enquote {\bibinfo {title} {{Dark matter
  in (partially) composite Higgs models}},}\ }\href {\doibase
  10.1007/JHEP12(2018)088} {\bibfield  {journal} {\bibinfo  {journal} {JHEP}\
  }\textbf {\bibinfo {volume} {12}},\ \bibinfo {pages} {088} (\bibinfo {year}
  {2018})},\ \Eprint {http://arxiv.org/abs/1808.07515} {arXiv:1808.07515
  [hep-ph]} \BibitemShut {NoStop}%
\bibitem [{\citenamefont {Balkin}\ \emph {et~al.}(2018)\citenamefont {Balkin},
  \citenamefont {Ruhdorfer}, \citenamefont {Salvioni},\ and\ \citenamefont
  {Weiler}}]{Balkin:2018tma}%
  \BibitemOpen
  \bibfield  {author} {\bibinfo {author} {\bibfnamefont {Reuven}\ \bibnamefont
  {Balkin}}, \bibinfo {author} {\bibfnamefont {Maximilian}\ \bibnamefont
  {Ruhdorfer}}, \bibinfo {author} {\bibfnamefont {Ennio}\ \bibnamefont
  {Salvioni}}, \ and\ \bibinfo {author} {\bibfnamefont {Andreas}\ \bibnamefont
  {Weiler}},\ }\bibfield  {title} {\enquote {\bibinfo {title} {{Dark matter
  shifts away from direct detection}},}\ }\href {\doibase
  10.1088/1475-7516/2018/11/050} {\bibfield  {journal} {\bibinfo  {journal}
  {JCAP}\ }\textbf {\bibinfo {volume} {11}},\ \bibinfo {pages} {050} (\bibinfo
  {year} {2018})},\ \Eprint {http://arxiv.org/abs/1809.09106} {arXiv:1809.09106
  [hep-ph]} \BibitemShut {NoStop}%
\bibitem [{\citenamefont {Cacciapaglia}\ \emph {et~al.}(2019)\citenamefont
  {Cacciapaglia}, \citenamefont {Cai}, \citenamefont {Deandrea},\ and\
  \citenamefont {Kushwaha}}]{Cacciapaglia:2019ixa}%
  \BibitemOpen
  \bibfield  {author} {\bibinfo {author} {\bibfnamefont {Giacomo}\ \bibnamefont
  {Cacciapaglia}}, \bibinfo {author} {\bibfnamefont {Haiying}\ \bibnamefont
  {Cai}}, \bibinfo {author} {\bibfnamefont {Aldo}\ \bibnamefont {Deandrea}}, \
  and\ \bibinfo {author} {\bibfnamefont {Ashwani}\ \bibnamefont {Kushwaha}},\
  }\bibfield  {title} {\enquote {\bibinfo {title} {{Composite Higgs and Dark
  Matter Model in SU(6)/SO(6)}},}\ }\href {\doibase 10.1007/JHEP10(2019)035}
  {\bibfield  {journal} {\bibinfo  {journal} {JHEP}\ }\textbf {\bibinfo
  {volume} {10}},\ \bibinfo {pages} {035} (\bibinfo {year} {2019})},\ \Eprint
  {http://arxiv.org/abs/1904.09301} {arXiv:1904.09301 [hep-ph]} \BibitemShut
  {NoStop}%
\bibitem [{\citenamefont {Ramos}(2020)}]{Ramos:2019qqa}%
  \BibitemOpen
  \bibfield  {author} {\bibinfo {author} {\bibfnamefont {Maria}\ \bibnamefont
  {Ramos}},\ }\bibfield  {title} {\enquote {\bibinfo {title} {{Composite dark
  matter phenomenology in the presence of lighter degrees of freedom}},}\
  }\href {\doibase 10.1007/JHEP07(2020)128} {\bibfield  {journal} {\bibinfo
  {journal} {JHEP}\ }\textbf {\bibinfo {volume} {07}},\ \bibinfo {pages} {128}
  (\bibinfo {year} {2020})},\ \Eprint {http://arxiv.org/abs/1912.11061}
  {arXiv:1912.11061 [hep-ph]} \BibitemShut {NoStop}%
\bibitem [{\citenamefont {Cai}\ and\ \citenamefont
  {Cacciapaglia}(2021)}]{Cai:2020njb}%
  \BibitemOpen
  \bibfield  {author} {\bibinfo {author} {\bibfnamefont {Haiying}\ \bibnamefont
  {Cai}}\ and\ \bibinfo {author} {\bibfnamefont {Giacomo}\ \bibnamefont
  {Cacciapaglia}},\ }\bibfield  {title} {\enquote {\bibinfo {title} {{Singlet
  dark matter in the SU(6)/SO(6) composite Higgs model}},}\ }\href {\doibase
  10.1103/PhysRevD.103.055002} {\bibfield  {journal} {\bibinfo  {journal}
  {Phys. Rev. D}\ }\textbf {\bibinfo {volume} {103}},\ \bibinfo {pages}
  {055002} (\bibinfo {year} {2021})},\ \Eprint
  {http://arxiv.org/abs/2007.04338} {arXiv:2007.04338 [hep-ph]} \BibitemShut
  {NoStop}%
\bibitem [{\citenamefont {Garny}\ \emph {et~al.}(2016)\citenamefont {Garny},
  \citenamefont {Sandora},\ and\ \citenamefont {Sloth}}]{Garny:2015sjg}%
  \BibitemOpen
  \bibfield  {author} {\bibinfo {author} {\bibfnamefont {Mathias}\ \bibnamefont
  {Garny}}, \bibinfo {author} {\bibfnamefont {McCullen}\ \bibnamefont
  {Sandora}}, \ and\ \bibinfo {author} {\bibfnamefont {Martin~S.}\ \bibnamefont
  {Sloth}},\ }\bibfield  {title} {\enquote {\bibinfo {title} {{Planckian
  Interacting Massive Particles as Dark Matter}},}\ }\href {\doibase
  10.1103/PhysRevLett.116.101302} {\bibfield  {journal} {\bibinfo  {journal}
  {Phys. Rev. Lett.}\ }\textbf {\bibinfo {volume} {116}},\ \bibinfo {pages}
  {101302} (\bibinfo {year} {2016})},\ \Eprint
  {http://arxiv.org/abs/1511.03278} {arXiv:1511.03278 [hep-ph]} \BibitemShut
  {NoStop}%
\bibitem [{\citenamefont {Garny}\ \emph {et~al.}(2018)\citenamefont {Garny},
  \citenamefont {Palessandro}, \citenamefont {Sandora},\ and\ \citenamefont
  {Sloth}}]{Garny:2017kha}%
  \BibitemOpen
  \bibfield  {author} {\bibinfo {author} {\bibfnamefont {Mathias}\ \bibnamefont
  {Garny}}, \bibinfo {author} {\bibfnamefont {Andrea}\ \bibnamefont
  {Palessandro}}, \bibinfo {author} {\bibfnamefont {McCullen}\ \bibnamefont
  {Sandora}}, \ and\ \bibinfo {author} {\bibfnamefont {Martin~S.}\ \bibnamefont
  {Sloth}},\ }\bibfield  {title} {\enquote {\bibinfo {title} {{Theory and
  Phenomenology of Planckian Interacting Massive Particles as Dark Matter}},}\
  }\href {\doibase 10.1088/1475-7516/2018/02/027} {\bibfield  {journal}
  {\bibinfo  {journal} {JCAP}\ }\textbf {\bibinfo {volume} {02}},\ \bibinfo
  {pages} {027} (\bibinfo {year} {2018})},\ \Eprint
  {http://arxiv.org/abs/1709.09688} {arXiv:1709.09688 [hep-ph]} \BibitemShut
  {NoStop}%
\bibitem [{\citenamefont {Chianese}\ \emph {et~al.}(2020)\citenamefont
  {Chianese}, \citenamefont {Fu},\ and\ \citenamefont
  {King}}]{Chianese:2019epo}%
  \BibitemOpen
  \bibfield  {author} {\bibinfo {author} {\bibfnamefont {Marco}\ \bibnamefont
  {Chianese}}, \bibinfo {author} {\bibfnamefont {Bowen}\ \bibnamefont {Fu}}, \
  and\ \bibinfo {author} {\bibfnamefont {Stephen~F.}\ \bibnamefont {King}},\
  }\bibfield  {title} {\enquote {\bibinfo {title} {{Minimal Seesaw extension
  for Neutrino Mass and Mixing, Leptogenesis and Dark Matter: FIMPzillas
  through the Right-Handed Neutrino Portal}},}\ }\href {\doibase
  10.1088/1475-7516/2020/03/030} {\bibfield  {journal} {\bibinfo  {journal}
  {JCAP}\ }\textbf {\bibinfo {volume} {03}},\ \bibinfo {pages} {030} (\bibinfo
  {year} {2020})},\ \Eprint {http://arxiv.org/abs/1910.12916} {arXiv:1910.12916
  [hep-ph]} \BibitemShut {NoStop}%
\bibitem [{\citenamefont {Georgi}(2007)}]{Georgi:2007ek}%
  \BibitemOpen
  \bibfield  {author} {\bibinfo {author} {\bibfnamefont {Howard}\ \bibnamefont
  {Georgi}},\ }\bibfield  {title} {\enquote {\bibinfo {title} {{Unparticle
  physics}},}\ }\href {\doibase 10.1103/PhysRevLett.98.221601} {\bibfield
  {journal} {\bibinfo  {journal} {Phys. Rev. Lett.}\ }\textbf {\bibinfo
  {volume} {98}},\ \bibinfo {pages} {221601} (\bibinfo {year} {2007})},\
  \Eprint {http://arxiv.org/abs/hep-ph/0703260} {arXiv:hep-ph/0703260}
  \BibitemShut {NoStop}%
\bibitem [{\citenamefont {Agashe}\ \emph {et~al.}(2005)\citenamefont {Agashe},
  \citenamefont {Contino},\ and\ \citenamefont {Pomarol}}]{Agashe:2004rs}%
  \BibitemOpen
  \bibfield  {author} {\bibinfo {author} {\bibfnamefont {Kaustubh}\
  \bibnamefont {Agashe}}, \bibinfo {author} {\bibfnamefont {Roberto}\
  \bibnamefont {Contino}}, \ and\ \bibinfo {author} {\bibfnamefont {Alex}\
  \bibnamefont {Pomarol}},\ }\bibfield  {title} {\enquote {\bibinfo {title}
  {{The Minimal composite Higgs model}},}\ }\href {\doibase
  10.1016/j.nuclphysb.2005.04.035} {\bibfield  {journal} {\bibinfo  {journal}
  {Nucl. Phys. B}\ }\textbf {\bibinfo {volume} {719}},\ \bibinfo {pages}
  {165--187} (\bibinfo {year} {2005})},\ \Eprint
  {http://arxiv.org/abs/hep-ph/0412089} {arXiv:hep-ph/0412089} \BibitemShut
  {NoStop}%
\bibitem [{\citenamefont {Agashe}\ \emph {et~al.}(2003)\citenamefont {Agashe},
  \citenamefont {Delgado}, \citenamefont {May},\ and\ \citenamefont
  {Sundrum}}]{Agashe:2003zs}%
  \BibitemOpen
  \bibfield  {author} {\bibinfo {author} {\bibfnamefont {Kaustubh}\
  \bibnamefont {Agashe}}, \bibinfo {author} {\bibfnamefont {Antonio}\
  \bibnamefont {Delgado}}, \bibinfo {author} {\bibfnamefont {Michael~J.}\
  \bibnamefont {May}}, \ and\ \bibinfo {author} {\bibfnamefont {Raman}\
  \bibnamefont {Sundrum}},\ }\bibfield  {title} {\enquote {\bibinfo {title}
  {{RS1, custodial isospin and precision tests}},}\ }\href {\doibase
  10.1088/1126-6708/2003/08/050} {\bibfield  {journal} {\bibinfo  {journal}
  {JHEP}\ }\textbf {\bibinfo {volume} {08}},\ \bibinfo {pages} {050} (\bibinfo
  {year} {2003})},\ \Eprint {http://arxiv.org/abs/hep-ph/0308036}
  {arXiv:hep-ph/0308036} \BibitemShut {NoStop}%
\bibitem [{\citenamefont {von Harling}\ and\ \citenamefont
  {Servant}(2018)}]{vonHarling:2017yew}%
  \BibitemOpen
  \bibfield  {author} {\bibinfo {author} {\bibfnamefont {Benedict}\
  \bibnamefont {von Harling}}\ and\ \bibinfo {author} {\bibfnamefont
  {Geraldine}\ \bibnamefont {Servant}},\ }\bibfield  {title} {\enquote
  {\bibinfo {title} {{QCD-induced Electroweak Phase Transition}},}\ }\href
  {\doibase 10.1007/JHEP01(2018)159} {\bibfield  {journal} {\bibinfo  {journal}
  {JHEP}\ }\textbf {\bibinfo {volume} {01}},\ \bibinfo {pages} {159} (\bibinfo
  {year} {2018})},\ \Eprint {http://arxiv.org/abs/1711.11554} {arXiv:1711.11554
  [hep-ph]} \BibitemShut {NoStop}%
\bibitem [{\citenamefont {Choi}\ and\ \citenamefont {Im}(2016)}]{Choi:2015fiu}%
  \BibitemOpen
  \bibfield  {author} {\bibinfo {author} {\bibfnamefont {Kiwoon}\ \bibnamefont
  {Choi}}\ and\ \bibinfo {author} {\bibfnamefont {Sang~Hui}\ \bibnamefont
  {Im}},\ }\bibfield  {title} {\enquote {\bibinfo {title} {{Realizing the
  relaxion from multiple axions and its UV completion with high scale
  supersymmetry}},}\ }\href {\doibase 10.1007/JHEP01(2016)149} {\bibfield
  {journal} {\bibinfo  {journal} {JHEP}\ }\textbf {\bibinfo {volume} {01}},\
  \bibinfo {pages} {149} (\bibinfo {year} {2016})},\ \Eprint
  {http://arxiv.org/abs/1511.00132} {arXiv:1511.00132 [hep-ph]} \BibitemShut
  {NoStop}%
\bibitem [{\citenamefont {Kaplan}\ and\ \citenamefont
  {Rattazzi}(2016)}]{Kaplan:2015fuy}%
  \BibitemOpen
  \bibfield  {author} {\bibinfo {author} {\bibfnamefont {David~E.}\
  \bibnamefont {Kaplan}}\ and\ \bibinfo {author} {\bibfnamefont {Riccardo}\
  \bibnamefont {Rattazzi}},\ }\bibfield  {title} {\enquote {\bibinfo {title}
  {{Large field excursions and approximate discrete symmetries from a clockwork
  axion}},}\ }\href {\doibase 10.1103/PhysRevD.93.085007} {\bibfield  {journal}
  {\bibinfo  {journal} {Phys. Rev. D}\ }\textbf {\bibinfo {volume} {93}},\
  \bibinfo {pages} {085007} (\bibinfo {year} {2016})},\ \Eprint
  {http://arxiv.org/abs/1511.01827} {arXiv:1511.01827 [hep-ph]} \BibitemShut
  {NoStop}%
\bibitem [{\citenamefont {Giudice}\ and\ \citenamefont
  {McCullough}(2017)}]{Giudice:2016yja}%
  \BibitemOpen
  \bibfield  {author} {\bibinfo {author} {\bibfnamefont {Gian~F.}\ \bibnamefont
  {Giudice}}\ and\ \bibinfo {author} {\bibfnamefont {Matthew}\ \bibnamefont
  {McCullough}},\ }\bibfield  {title} {\enquote {\bibinfo {title} {{A Clockwork
  Theory}},}\ }\href {\doibase 10.1007/JHEP02(2017)036} {\bibfield  {journal}
  {\bibinfo  {journal} {JHEP}\ }\textbf {\bibinfo {volume} {02}},\ \bibinfo
  {pages} {036} (\bibinfo {year} {2017})},\ \Eprint
  {http://arxiv.org/abs/1610.07962} {arXiv:1610.07962 [hep-ph]} \BibitemShut
  {NoStop}%
\bibitem [{\citenamefont {Silverstein}\ and\ \citenamefont
  {Westphal}(2008)}]{Silverstein:2008sg}%
  \BibitemOpen
  \bibfield  {author} {\bibinfo {author} {\bibfnamefont {Eva}\ \bibnamefont
  {Silverstein}}\ and\ \bibinfo {author} {\bibfnamefont {Alexander}\
  \bibnamefont {Westphal}},\ }\bibfield  {title} {\enquote {\bibinfo {title}
  {{Monodromy in the CMB: Gravity Waves and String Inflation}},}\ }\href
  {\doibase 10.1103/PhysRevD.78.106003} {\bibfield  {journal} {\bibinfo
  {journal} {Phys. Rev. D}\ }\textbf {\bibinfo {volume} {78}},\ \bibinfo
  {pages} {106003} (\bibinfo {year} {2008})},\ \Eprint
  {http://arxiv.org/abs/0803.3085} {arXiv:0803.3085 [hep-th]} \BibitemShut
  {NoStop}%
\bibitem [{\citenamefont {Redi}\ and\ \citenamefont
  {Tesi}(2021)}]{Redi:2021ipn}%
  \BibitemOpen
  \bibfield  {author} {\bibinfo {author} {\bibfnamefont {Michele}\ \bibnamefont
  {Redi}}\ and\ \bibinfo {author} {\bibfnamefont {Andrea}\ \bibnamefont
  {Tesi}},\ }\bibfield  {title} {\enquote {\bibinfo {title} {{General freeze-in
  and freeze-out}},}\ }\href {\doibase 10.1007/JHEP12(2021)060} {\bibfield
  {journal} {\bibinfo  {journal} {JHEP}\ }\textbf {\bibinfo {volume} {12}},\
  \bibinfo {pages} {060} (\bibinfo {year} {2021})},\ \Eprint
  {http://arxiv.org/abs/2107.14801} {arXiv:2107.14801 [hep-ph]} \BibitemShut
  {NoStop}%
\bibitem [{\citenamefont {Aghanim}\ \emph {et~al.}(2020)\citenamefont {Aghanim}
  \emph {et~al.}}]{Planck:2018vyg}%
  \BibitemOpen
  \bibfield  {author} {\bibinfo {author} {\bibfnamefont {N.}~\bibnamefont
  {Aghanim}} \emph {et~al.} (\bibinfo {collaboration} {Planck}),\ }\bibfield
  {title} {\enquote {\bibinfo {title} {{Planck 2018 results. VI. Cosmological
  parameters}},}\ }\href {\doibase 10.1051/0004-6361/201833910} {\bibfield
  {journal} {\bibinfo  {journal} {Astron. Astrophys.}\ }\textbf {\bibinfo
  {volume} {641}},\ \bibinfo {pages} {A6} (\bibinfo {year} {2020})},\ \bibinfo
  {note} {[Erratum: Astron.Astrophys. 652, C4 (2021)]},\ \Eprint
  {http://arxiv.org/abs/1807.06209} {arXiv:1807.06209 [astro-ph.CO]}
  \BibitemShut {NoStop}%
\bibitem [{\citenamefont {Workman}\ \emph {et~al.}(2022)\citenamefont {Workman}
  \emph {et~al.}}]{ParticleDataGroup:2022pth}%
  \BibitemOpen
  \bibfield  {author} {\bibinfo {author} {\bibfnamefont {R.~L.}\ \bibnamefont
  {Workman}} \emph {et~al.} (\bibinfo {collaboration} {Particle Data Group}),\
  }\bibfield  {title} {\enquote {\bibinfo {title} {{Review of Particle
  Physics}},}\ }\href {\doibase 10.1093/ptep/ptac097} {\bibfield  {journal}
  {\bibinfo  {journal} {PTEP}\ }\textbf {\bibinfo {volume} {2022}},\ \bibinfo
  {pages} {083C01} (\bibinfo {year} {2022})}\BibitemShut {NoStop}%
\bibitem [{\citenamefont {Ade}\ \emph {et~al.}(2021)\citenamefont {Ade} \emph
  {et~al.}}]{BICEP:2021xfz}%
  \BibitemOpen
  \bibfield  {author} {\bibinfo {author} {\bibfnamefont {P.~A.~R.}\
  \bibnamefont {Ade}} \emph {et~al.} (\bibinfo {collaboration} {BICEP, Keck}),\
  }\bibfield  {title} {\enquote {\bibinfo {title} {{Improved Constraints on
  Primordial Gravitational Waves using Planck, WMAP, and BICEP/Keck
  Observations through the 2018 Observing Season}},}\ }\href {\doibase
  10.1103/PhysRevLett.127.151301} {\bibfield  {journal} {\bibinfo  {journal}
  {Phys. Rev. Lett.}\ }\textbf {\bibinfo {volume} {127}},\ \bibinfo {pages}
  {151301} (\bibinfo {year} {2021})},\ \Eprint
  {http://arxiv.org/abs/2110.00483} {arXiv:2110.00483 [astro-ph.CO]}
  \BibitemShut {NoStop}%
\bibitem [{\citenamefont {Akrami}\ \emph {et~al.}(2020)\citenamefont {Akrami}
  \emph {et~al.}}]{Planck:2018jri}%
  \BibitemOpen
  \bibfield  {author} {\bibinfo {author} {\bibfnamefont {Y.}~\bibnamefont
  {Akrami}} \emph {et~al.} (\bibinfo {collaboration} {Planck}),\ }\bibfield
  {title} {\enquote {\bibinfo {title} {{Planck 2018 results. X. Constraints on
  inflation}},}\ }\href {\doibase 10.1051/0004-6361/201833887} {\bibfield
  {journal} {\bibinfo  {journal} {Astron. Astrophys.}\ }\textbf {\bibinfo
  {volume} {641}},\ \bibinfo {pages} {A10} (\bibinfo {year} {2020})},\ \Eprint
  {http://arxiv.org/abs/1807.06211} {arXiv:1807.06211 [astro-ph.CO]}
  \BibitemShut {NoStop}%
\bibitem [{\citenamefont {Abazajian}\ \emph {et~al.}(2016)\citenamefont
  {Abazajian} \emph {et~al.}}]{CMB-S4:2016ple}%
  \BibitemOpen
  \bibfield  {author} {\bibinfo {author} {\bibfnamefont {Kevork~N.}\
  \bibnamefont {Abazajian}} \emph {et~al.} (\bibinfo {collaboration}
  {CMB-S4}),\ }\bibfield  {title} {\enquote {\bibinfo {title} {{CMB-S4 Science
  Book, First Edition}},}\ }\href@noop {} {\  (\bibinfo {year} {2016})},\
  \Eprint {http://arxiv.org/abs/1610.02743} {arXiv:1610.02743 [astro-ph.CO]}
  \BibitemShut {NoStop}%
\bibitem [{\citenamefont {Arkani-Hamed}\ and\ \citenamefont
  {Maldacena}(2015)}]{Arkani-Hamed:2015bza}%
  \BibitemOpen
  \bibfield  {author} {\bibinfo {author} {\bibfnamefont {Nima}\ \bibnamefont
  {Arkani-Hamed}}\ and\ \bibinfo {author} {\bibfnamefont {Juan}\ \bibnamefont
  {Maldacena}},\ }\bibfield  {title} {\enquote {\bibinfo {title} {{Cosmological
  Collider Physics}},}\ }\href@noop {} {\  (\bibinfo {year} {2015})},\ \Eprint
  {http://arxiv.org/abs/1503.08043} {arXiv:1503.08043 [hep-th]} \BibitemShut
  {NoStop}%
\bibitem [{\citenamefont {Carney}\ \emph {et~al.}(2020)\citenamefont {Carney},
  \citenamefont {Ghosh}, \citenamefont {Krnjaic},\ and\ \citenamefont
  {Taylor}}]{Carney:2019pza}%
  \BibitemOpen
  \bibfield  {author} {\bibinfo {author} {\bibfnamefont {Daniel}\ \bibnamefont
  {Carney}}, \bibinfo {author} {\bibfnamefont {Sohitri}\ \bibnamefont {Ghosh}},
  \bibinfo {author} {\bibfnamefont {Gordan}\ \bibnamefont {Krnjaic}}, \ and\
  \bibinfo {author} {\bibfnamefont {Jacob~M.}\ \bibnamefont {Taylor}},\
  }\bibfield  {title} {\enquote {\bibinfo {title} {{Proposal for gravitational
  direct detection of dark matter}},}\ }\href {\doibase
  10.1103/PhysRevD.102.072003} {\bibfield  {journal} {\bibinfo  {journal}
  {Phys. Rev. D}\ }\textbf {\bibinfo {volume} {102}},\ \bibinfo {pages}
  {072003} (\bibinfo {year} {2020})},\ \Eprint
  {http://arxiv.org/abs/1903.00492} {arXiv:1903.00492 [hep-ph]} \BibitemShut
  {NoStop}%
\bibitem [{\citenamefont {Attanasio}\ \emph {et~al.}(2022)\citenamefont
  {Attanasio} \emph {et~al.}}]{Windchime:2022whs}%
  \BibitemOpen
  \bibfield  {author} {\bibinfo {author} {\bibfnamefont {Alaina}\ \bibnamefont
  {Attanasio}} \emph {et~al.} (\bibinfo {collaboration} {Windchime}),\
  }\bibfield  {title} {\enquote {\bibinfo {title} {{Snowmass 2021 White Paper:
  The Windchime Project}},}\ }in\ \href@noop {} {\emph {\bibinfo {booktitle}
  {{2022 Snowmass Summer Study}}}}\ (\bibinfo {year} {2022})\ \Eprint
  {http://arxiv.org/abs/2203.07242} {arXiv:2203.07242 [hep-ex]} \BibitemShut
  {NoStop}%
\bibitem [{\citenamefont {Goldberger}\ and\ \citenamefont
  {Wise}(1999)}]{Goldberger:1999uk}%
  \BibitemOpen
  \bibfield  {author} {\bibinfo {author} {\bibfnamefont {Walter~D.}\
  \bibnamefont {Goldberger}}\ and\ \bibinfo {author} {\bibfnamefont {Mark~B.}\
  \bibnamefont {Wise}},\ }\bibfield  {title} {\enquote {\bibinfo {title}
  {{Modulus stabilization with bulk fields}},}\ }\href {\doibase
  10.1103/PhysRevLett.83.4922} {\bibfield  {journal} {\bibinfo  {journal}
  {Phys. Rev. Lett.}\ }\textbf {\bibinfo {volume} {83}},\ \bibinfo {pages}
  {4922--4925} (\bibinfo {year} {1999})},\ \Eprint
  {http://arxiv.org/abs/hep-ph/9907447} {arXiv:hep-ph/9907447} \BibitemShut
  {NoStop}%
\bibitem [{\citenamefont {Creminelli}\ \emph {et~al.}(2002)\citenamefont
  {Creminelli}, \citenamefont {Nicolis},\ and\ \citenamefont
  {Rattazzi}}]{Creminelli:2001th}%
  \BibitemOpen
  \bibfield  {author} {\bibinfo {author} {\bibfnamefont {Paolo}\ \bibnamefont
  {Creminelli}}, \bibinfo {author} {\bibfnamefont {Alberto}\ \bibnamefont
  {Nicolis}}, \ and\ \bibinfo {author} {\bibfnamefont {Riccardo}\ \bibnamefont
  {Rattazzi}},\ }\bibfield  {title} {\enquote {\bibinfo {title} {{Holography
  and the electroweak phase transition}},}\ }\href {\doibase
  10.1088/1126-6708/2002/03/051} {\bibfield  {journal} {\bibinfo  {journal}
  {JHEP}\ }\textbf {\bibinfo {volume} {03}},\ \bibinfo {pages} {051} (\bibinfo
  {year} {2002})},\ \Eprint {http://arxiv.org/abs/hep-th/0107141}
  {arXiv:hep-th/0107141} \BibitemShut {NoStop}%
\bibitem [{\citenamefont {Randall}\ and\ \citenamefont
  {Servant}(2007)}]{Randall:2006py}%
  \BibitemOpen
  \bibfield  {author} {\bibinfo {author} {\bibfnamefont {Lisa}\ \bibnamefont
  {Randall}}\ and\ \bibinfo {author} {\bibfnamefont {Geraldine}\ \bibnamefont
  {Servant}},\ }\bibfield  {title} {\enquote {\bibinfo {title} {{Gravitational
  waves from warped spacetime}},}\ }\href {\doibase
  10.1088/1126-6708/2007/05/054} {\bibfield  {journal} {\bibinfo  {journal}
  {JHEP}\ }\textbf {\bibinfo {volume} {05}},\ \bibinfo {pages} {054} (\bibinfo
  {year} {2007})},\ \Eprint {http://arxiv.org/abs/hep-ph/0607158}
  {arXiv:hep-ph/0607158} \BibitemShut {NoStop}%
\bibitem [{\citenamefont {Konstandin}\ and\ \citenamefont
  {Servant}(2011)}]{Konstandin:2011dr}%
  \BibitemOpen
  \bibfield  {author} {\bibinfo {author} {\bibfnamefont {Thomas}\ \bibnamefont
  {Konstandin}}\ and\ \bibinfo {author} {\bibfnamefont {Geraldine}\
  \bibnamefont {Servant}},\ }\bibfield  {title} {\enquote {\bibinfo {title}
  {{Cosmological Consequences of Nearly Conformal Dynamics at the TeV
  scale}},}\ }\href {\doibase 10.1088/1475-7516/2011/12/009} {\bibfield
  {journal} {\bibinfo  {journal} {JCAP}\ }\textbf {\bibinfo {volume} {12}},\
  \bibinfo {pages} {009} (\bibinfo {year} {2011})},\ \Eprint
  {http://arxiv.org/abs/1104.4791} {arXiv:1104.4791 [hep-ph]} \BibitemShut
  {NoStop}%
\bibitem [{\citenamefont {Baratella}\ \emph {et~al.}(2019)\citenamefont
  {Baratella}, \citenamefont {Pomarol},\ and\ \citenamefont
  {Rompineve}}]{Baratella:2018pxi}%
  \BibitemOpen
  \bibfield  {author} {\bibinfo {author} {\bibfnamefont {Pietro}\ \bibnamefont
  {Baratella}}, \bibinfo {author} {\bibfnamefont {Alex}\ \bibnamefont
  {Pomarol}}, \ and\ \bibinfo {author} {\bibfnamefont {Fabrizio}\ \bibnamefont
  {Rompineve}},\ }\bibfield  {title} {\enquote {\bibinfo {title} {{The
  Supercooled Universe}},}\ }\href {\doibase 10.1007/JHEP03(2019)100}
  {\bibfield  {journal} {\bibinfo  {journal} {JHEP}\ }\textbf {\bibinfo
  {volume} {03}},\ \bibinfo {pages} {100} (\bibinfo {year} {2019})},\ \Eprint
  {http://arxiv.org/abs/1812.06996} {arXiv:1812.06996 [hep-ph]} \BibitemShut
  {NoStop}%
\bibitem [{\citenamefont {Hassanain}\ \emph {et~al.}(2007)\citenamefont
  {Hassanain}, \citenamefont {March-Russell},\ and\ \citenamefont
  {Schvellinger}}]{Hassanain:2007js}%
  \BibitemOpen
  \bibfield  {author} {\bibinfo {author} {\bibfnamefont {Babiker}\ \bibnamefont
  {Hassanain}}, \bibinfo {author} {\bibfnamefont {John}\ \bibnamefont
  {March-Russell}}, \ and\ \bibinfo {author} {\bibfnamefont {Martin}\
  \bibnamefont {Schvellinger}},\ }\bibfield  {title} {\enquote {\bibinfo
  {title} {{Warped Deformed Throats have Faster (Electroweak) Phase
  Transitions}},}\ }\href {\doibase 10.1088/1126-6708/2007/10/089} {\bibfield
  {journal} {\bibinfo  {journal} {JHEP}\ }\textbf {\bibinfo {volume} {10}},\
  \bibinfo {pages} {089} (\bibinfo {year} {2007})},\ \Eprint
  {http://arxiv.org/abs/0708.2060} {arXiv:0708.2060 [hep-th]} \BibitemShut
  {NoStop}%
\bibitem [{\citenamefont {Bunk}\ \emph {et~al.}(2018)\citenamefont {Bunk},
  \citenamefont {Hubisz},\ and\ \citenamefont {Jain}}]{Bunk:2017fic}%
  \BibitemOpen
  \bibfield  {author} {\bibinfo {author} {\bibfnamefont {Don}\ \bibnamefont
  {Bunk}}, \bibinfo {author} {\bibfnamefont {Jay}\ \bibnamefont {Hubisz}}, \
  and\ \bibinfo {author} {\bibfnamefont {Bithika}\ \bibnamefont {Jain}},\
  }\bibfield  {title} {\enquote {\bibinfo {title} {{A Perturbative RS I
  Cosmological Phase Transition}},}\ }\href {\doibase
  10.1140/epjc/s10052-018-5529-2} {\bibfield  {journal} {\bibinfo  {journal}
  {Eur. Phys. J. C}\ }\textbf {\bibinfo {volume} {78}},\ \bibinfo {pages} {78}
  (\bibinfo {year} {2018})},\ \Eprint {http://arxiv.org/abs/1705.00001}
  {arXiv:1705.00001 [hep-ph]} \BibitemShut {NoStop}%
\bibitem [{\citenamefont {Dillon}\ \emph {et~al.}(2018)\citenamefont {Dillon},
  \citenamefont {El-Menoufi}, \citenamefont {Huber},\ and\ \citenamefont
  {Manuel}}]{Dillon:2017ctw}%
  \BibitemOpen
  \bibfield  {author} {\bibinfo {author} {\bibfnamefont {Barry~M.}\
  \bibnamefont {Dillon}}, \bibinfo {author} {\bibfnamefont {Basem~Kamal}\
  \bibnamefont {El-Menoufi}}, \bibinfo {author} {\bibfnamefont {Stephan~J.}\
  \bibnamefont {Huber}}, \ and\ \bibinfo {author} {\bibfnamefont {Jonathan~P.}\
  \bibnamefont {Manuel}},\ }\bibfield  {title} {\enquote {\bibinfo {title}
  {{Rapid holographic phase transition with brane-localized curvature}},}\
  }\href {\doibase 10.1103/PhysRevD.98.086005} {\bibfield  {journal} {\bibinfo
  {journal} {Phys. Rev. D}\ }\textbf {\bibinfo {volume} {98}},\ \bibinfo
  {pages} {086005} (\bibinfo {year} {2018})},\ \Eprint
  {http://arxiv.org/abs/1708.02953} {arXiv:1708.02953 [hep-th]} \BibitemShut
  {NoStop}%
\bibitem [{\citenamefont {Meg\'\i{}as}\ \emph {et~al.}(2018)\citenamefont
  {Meg\'\i{}as}, \citenamefont {Nardini},\ and\ \citenamefont
  {Quir\'os}}]{Megias:2018sxv}%
  \BibitemOpen
  \bibfield  {author} {\bibinfo {author} {\bibfnamefont {Eugenio}\ \bibnamefont
  {Meg\'\i{}as}}, \bibinfo {author} {\bibfnamefont {Germano}\ \bibnamefont
  {Nardini}}, \ and\ \bibinfo {author} {\bibfnamefont {Mariano}\ \bibnamefont
  {Quir\'os}},\ }\bibfield  {title} {\enquote {\bibinfo {title} {{Cosmological
  Phase Transitions in Warped Space: Gravitational Waves and Collider
  Signatures}},}\ }\href {\doibase 10.1007/JHEP09(2018)095} {\bibfield
  {journal} {\bibinfo  {journal} {JHEP}\ }\textbf {\bibinfo {volume} {09}},\
  \bibinfo {pages} {095} (\bibinfo {year} {2018})},\ \Eprint
  {http://arxiv.org/abs/1806.04877} {arXiv:1806.04877 [hep-ph]} \BibitemShut
  {NoStop}%
\bibitem [{\citenamefont {Agashe}\ \emph {et~al.}(2020)\citenamefont {Agashe},
  \citenamefont {Du}, \citenamefont {Ekhterachian}, \citenamefont {Kumar},\
  and\ \citenamefont {Sundrum}}]{Agashe:2019lhy}%
  \BibitemOpen
  \bibfield  {author} {\bibinfo {author} {\bibfnamefont {Kaustubh}\
  \bibnamefont {Agashe}}, \bibinfo {author} {\bibfnamefont {Peizhi}\
  \bibnamefont {Du}}, \bibinfo {author} {\bibfnamefont {Majid}\ \bibnamefont
  {Ekhterachian}}, \bibinfo {author} {\bibfnamefont {Soubhik}\ \bibnamefont
  {Kumar}}, \ and\ \bibinfo {author} {\bibfnamefont {Raman}\ \bibnamefont
  {Sundrum}},\ }\bibfield  {title} {\enquote {\bibinfo {title} {{Cosmological
  Phase Transition of Spontaneous Confinement}},}\ }\href {\doibase
  10.1007/JHEP05(2020)086} {\bibfield  {journal} {\bibinfo  {journal} {JHEP}\
  }\textbf {\bibinfo {volume} {05}},\ \bibinfo {pages} {086} (\bibinfo {year}
  {2020})},\ \Eprint {http://arxiv.org/abs/1910.06238} {arXiv:1910.06238
  [hep-ph]} \BibitemShut {NoStop}%
\bibitem [{\citenamefont {Cs\'aki}\ \emph {et~al.}(2023)\citenamefont
  {Cs\'aki}, \citenamefont {Geller}, \citenamefont {Heller-Algazi},\ and\
  \citenamefont {Ismail}}]{Csaki:2023pwy}%
  \BibitemOpen
  \bibfield  {author} {\bibinfo {author} {\bibfnamefont {Csaba}\ \bibnamefont
  {Cs\'aki}}, \bibinfo {author} {\bibfnamefont {Michael}\ \bibnamefont
  {Geller}}, \bibinfo {author} {\bibfnamefont {Zamir}\ \bibnamefont
  {Heller-Algazi}}, \ and\ \bibinfo {author} {\bibfnamefont {Ameen}\
  \bibnamefont {Ismail}},\ }\bibfield  {title} {\enquote {\bibinfo {title}
  {{Relevant Dilaton Stabilization}},}\ }\href@noop {} {\  (\bibinfo {year}
  {2023})},\ \Eprint {http://arxiv.org/abs/2301.10247} {arXiv:2301.10247
  [hep-ph]} \BibitemShut {NoStop}%
\bibitem [{\citenamefont {Er\"oncel}\ \emph {et~al.}(2023)\citenamefont
  {Er\"oncel}, \citenamefont {Hubisz}, \citenamefont {Lee}, \citenamefont
  {Rigo},\ and\ \citenamefont {Sambasivam}}]{Eroncel:2023uqf}%
  \BibitemOpen
  \bibfield  {author} {\bibinfo {author} {\bibfnamefont {Cem}\ \bibnamefont
  {Er\"oncel}}, \bibinfo {author} {\bibfnamefont {Jay}\ \bibnamefont {Hubisz}},
  \bibinfo {author} {\bibfnamefont {Seung.~J.}\ \bibnamefont {Lee}}, \bibinfo
  {author} {\bibfnamefont {Gabriele}\ \bibnamefont {Rigo}}, \ and\ \bibinfo
  {author} {\bibfnamefont {Bharath}\ \bibnamefont {Sambasivam}},\ }\bibfield
  {title} {\enquote {\bibinfo {title} {{New Horizons in the Holographic
  Conformal Phase Transition}},}\ }\href@noop {} {\  (\bibinfo {year}
  {2023})},\ \Eprint {http://arxiv.org/abs/2305.03773} {arXiv:2305.03773
  [hep-ph]} \BibitemShut {NoStop}%
\bibitem [{\citenamefont {Agrawal}\ and\ \citenamefont
  {Nee}(2021)}]{Agrawal:2021alq}%
  \BibitemOpen
  \bibfield  {author} {\bibinfo {author} {\bibfnamefont {Prateek}\ \bibnamefont
  {Agrawal}}\ and\ \bibinfo {author} {\bibfnamefont {Michael}\ \bibnamefont
  {Nee}},\ }\bibfield  {title} {\enquote {\bibinfo {title} {{Avoided
  deconfinement in Randall-Sundrum models}},}\ }\href {\doibase
  10.1007/JHEP10(2021)105} {\bibfield  {journal} {\bibinfo  {journal} {JHEP}\
  }\textbf {\bibinfo {volume} {10}},\ \bibinfo {pages} {105} (\bibinfo {year}
  {2021})},\ \Eprint {http://arxiv.org/abs/2103.05646} {arXiv:2103.05646
  [hep-ph]} \BibitemShut {NoStop}%
\bibitem [{\citenamefont {Kachru}\ \emph {et~al.}(2003)\citenamefont {Kachru},
  \citenamefont {Kallosh}, \citenamefont {Linde}, \citenamefont {Maldacena},
  \citenamefont {McAllister},\ and\ \citenamefont {Trivedi}}]{Kachru:2003sx}%
  \BibitemOpen
  \bibfield  {author} {\bibinfo {author} {\bibfnamefont {Shamit}\ \bibnamefont
  {Kachru}}, \bibinfo {author} {\bibfnamefont {Renata}\ \bibnamefont
  {Kallosh}}, \bibinfo {author} {\bibfnamefont {Andrei~D.}\ \bibnamefont
  {Linde}}, \bibinfo {author} {\bibfnamefont {Juan~Martin}\ \bibnamefont
  {Maldacena}}, \bibinfo {author} {\bibfnamefont {Liam~P.}\ \bibnamefont
  {McAllister}}, \ and\ \bibinfo {author} {\bibfnamefont {Sandip~P.}\
  \bibnamefont {Trivedi}},\ }\bibfield  {title} {\enquote {\bibinfo {title}
  {{Towards inflation in string theory}},}\ }\href {\doibase
  10.1088/1475-7516/2003/10/013} {\bibfield  {journal} {\bibinfo  {journal}
  {JCAP}\ }\textbf {\bibinfo {volume} {10}},\ \bibinfo {pages} {013} (\bibinfo
  {year} {2003})},\ \Eprint {http://arxiv.org/abs/hep-th/0308055}
  {arXiv:hep-th/0308055} \BibitemShut {NoStop}%
\bibitem [{\citenamefont {Bassett}\ \emph {et~al.}(2006)\citenamefont
  {Bassett}, \citenamefont {Tsujikawa},\ and\ \citenamefont
  {Wands}}]{Bassett:2005xm}%
  \BibitemOpen
  \bibfield  {author} {\bibinfo {author} {\bibfnamefont {Bruce~A.}\
  \bibnamefont {Bassett}}, \bibinfo {author} {\bibfnamefont {Shinji}\
  \bibnamefont {Tsujikawa}}, \ and\ \bibinfo {author} {\bibfnamefont {David}\
  \bibnamefont {Wands}},\ }\bibfield  {title} {\enquote {\bibinfo {title}
  {{Inflation dynamics and reheating}},}\ }\href {\doibase
  10.1103/RevModPhys.78.537} {\bibfield  {journal} {\bibinfo  {journal} {Rev.
  Mod. Phys.}\ }\textbf {\bibinfo {volume} {78}},\ \bibinfo {pages} {537--589}
  (\bibinfo {year} {2006})},\ \Eprint {http://arxiv.org/abs/astro-ph/0507632}
  {arXiv:astro-ph/0507632} \BibitemShut {NoStop}%
\bibitem [{\citenamefont {Douglas}\ and\ \citenamefont
  {Kachru}(2007)}]{Douglas:2006es}%
  \BibitemOpen
  \bibfield  {author} {\bibinfo {author} {\bibfnamefont {Michael~R.}\
  \bibnamefont {Douglas}}\ and\ \bibinfo {author} {\bibfnamefont {Shamit}\
  \bibnamefont {Kachru}},\ }\bibfield  {title} {\enquote {\bibinfo {title}
  {{Flux compactification}},}\ }\href {\doibase 10.1103/RevModPhys.79.733}
  {\bibfield  {journal} {\bibinfo  {journal} {Rev. Mod. Phys.}\ }\textbf
  {\bibinfo {volume} {79}},\ \bibinfo {pages} {733--796} (\bibinfo {year}
  {2007})},\ \Eprint {http://arxiv.org/abs/hep-th/0610102}
  {arXiv:hep-th/0610102} \BibitemShut {NoStop}%
\bibitem [{\citenamefont {McAllister}\ \emph {et~al.}(2010)\citenamefont
  {McAllister}, \citenamefont {Silverstein},\ and\ \citenamefont
  {Westphal}}]{McAllister:2008hb}%
  \BibitemOpen
  \bibfield  {author} {\bibinfo {author} {\bibfnamefont {Liam}\ \bibnamefont
  {McAllister}}, \bibinfo {author} {\bibfnamefont {Eva}\ \bibnamefont
  {Silverstein}}, \ and\ \bibinfo {author} {\bibfnamefont {Alexander}\
  \bibnamefont {Westphal}},\ }\bibfield  {title} {\enquote {\bibinfo {title}
  {{Gravity Waves and Linear Inflation from Axion Monodromy}},}\ }\href
  {\doibase 10.1103/PhysRevD.82.046003} {\bibfield  {journal} {\bibinfo
  {journal} {Phys. Rev. D}\ }\textbf {\bibinfo {volume} {82}},\ \bibinfo
  {pages} {046003} (\bibinfo {year} {2010})},\ \Eprint
  {http://arxiv.org/abs/0808.0706} {arXiv:0808.0706 [hep-th]} \BibitemShut
  {NoStop}%
\bibitem [{\citenamefont {Baumann}\ and\ \citenamefont
  {McAllister}(2015)}]{Baumann:2014nda}%
  \BibitemOpen
  \bibfield  {author} {\bibinfo {author} {\bibfnamefont {Daniel}\ \bibnamefont
  {Baumann}}\ and\ \bibinfo {author} {\bibfnamefont {Liam}\ \bibnamefont
  {McAllister}},\ }\href {\doibase 10.1017/CBO9781316105733} {\emph {\bibinfo
  {title} {{Inflation and String Theory}}}},\ Cambridge Monographs on
  Mathematical Physics\ (\bibinfo  {publisher} {Cambridge University Press},\
  \bibinfo {year} {2015})\ \Eprint {http://arxiv.org/abs/1404.2601}
  {arXiv:1404.2601 [hep-th]} \BibitemShut {NoStop}%
\bibitem [{\citenamefont {Gondolo}\ and\ \citenamefont
  {Gelmini}(1991)}]{Gondolo:1990dk}%
  \BibitemOpen
  \bibfield  {author} {\bibinfo {author} {\bibfnamefont {Paolo}\ \bibnamefont
  {Gondolo}}\ and\ \bibinfo {author} {\bibfnamefont {Graciela}\ \bibnamefont
  {Gelmini}},\ }\bibfield  {title} {\enquote {\bibinfo {title} {{Cosmic
  abundances of stable particles: Improved analysis}},}\ }\href {\doibase
  10.1016/0550-3213(91)90438-4} {\bibfield  {journal} {\bibinfo  {journal}
  {Nucl. Phys. B}\ }\textbf {\bibinfo {volume} {360}},\ \bibinfo {pages}
  {145--179} (\bibinfo {year} {1991})}\BibitemShut {NoStop}%
\end{thebibliography}%

\end{document}